\documentclass[11pt]{article}
\pdfoutput=1
\usepackage{graphicx}
\usepackage{epsfig}
\usepackage{amsfonts}
\usepackage{amscd}
\usepackage{latexsym}
\usepackage{amsmath,amssymb}
\usepackage{verbatim}
\usepackage{setspace}
\usepackage{color}
\usepackage{fancyhdr}
\usepackage{cite}
\usepackage{hyperref}
\usepackage{tensor}
\usepackage{xfrac}

\usepackage[textheight=9in, textwidth=6.5in, letterpaper]{geometry}

\numberwithin{equation}{section}

\def\0{{(0)}}
\def\1{{(1)}}
\def\2{{(2)}}

\def\<{\langle }
\def\>{\rangle }

\newcommand{\bea}{\begin{eqnarray}}
\newcommand{\eea}{\end{eqnarray}}
\newcommand{\be}{\begin{equation}}
\newcommand{\ee}{\end{equation}}
\newcommand{\ba}{\begin{aligned}}
\newcommand{\ea}{\end{aligned}}

\def\be{\begin{equation}}
\def\ee{\end{equation}}
\def\beq{\be\begin{array}{c}}
\def\eeq{\end{array}\ee}

   \makeatletter
  \let\over=\@@over \let\overwithdelims=\@@overwithdelims
  \let\atop=\@@atop \let\atopwithdelims=\@@atopwithdelims
  \let\above=\@@above \let\abovewithdelims=\@@abovewithdelims
\renewcommand\section{\@startsection {section}{1}{\z@}%
                                   {-3.5ex \@plus -1ex \@minus -.2ex}
                                   {2.3ex \@plus.2ex}%
                                   {\normalfont\large\bfseries}}

\renewcommand\subsection{\@startsection{subsection}{2}{\z@}%
                                     {-3.25ex\@plus -1ex \@minus -.2ex}%
                                     {1.5ex \@plus .2ex}%
                                     {\normalfont\bfseries}}

\usepackage{tikz}
\usepackage{slashed}
\usetikzlibrary{calc}   
\usetikzlibrary{topaths}
\usetikzlibrary{decorations}
\usetikzlibrary{decorations.pathmorphing}

{\cfoot{\thepage}}
\begin{document}
\begin{titlepage}
\unitlength = 1mm~\\
\vskip 3cm
\begin{center}

{\LARGE{\textsc{Celestial OPE blocks}}}

\vspace{0.8cm}
Alfredo Guevara{$^{*\dagger\mathsection}$}\\
\vspace{1cm}

{$^*$\it  Center for the Fundamental Laws of Nature, Harvard University, Cambridge, MA 02138, USA} \\
{$^\dagger$\it Black Hole Initiative, Harvard University, Cambridge, MA 02138, USA}\\
{$^\mathsection$\it Society of Fellows, Harvard University, Cambridge, MA 02138, USA}

\vspace{0.8cm}

\begin{abstract}
Starting from the defining two-point and three-point functions of Celestial CFTs,
Euclidean integral blocks are constructed for the OPE of scalar primaries. In their integral form they can alternatively be fixed using Poincar\'e symmetry acting on both massless and massive states. Subsequently, an analytic continuation
is done to define the Lorentzian version of the correlation functions and the OPE blocks as valued on the $(1,1)$ cylinder, the universal cover of the recently studied
celestial torus. The continuation is essentially the same
that is used in the derivation of the KLT relations for string amplitudes.
It is shown explicitly that the continued OPE blocks encode
the contributions from massive primary states as well as their shadow and
light-transformed partners of continuous spin. The corresponding pairings are also studied, thus the construction provides the fundamental relation between correlation functions and OPE coefficients in the scalar CCFT.
\end{abstract}

\end{center}

\end{titlepage}

\tableofcontents

\section{Introduction}

In the quest for a formulation of flat holography, Celestial Conformal Field Theories (CCFTs) constitute a novel approach: By introducing conformal correlation functions living on the celestial sphere they provide a dual description of scattering in asymptotically flat spacetimes, see \cite{Pasterski:2021rjz,Raclariu:2021zjz} for recent reviews. In turn, these correlation functions are obtained from the S-matrix of a particular set of wavefunctions in the bulk \cite{deBoer:2003vf,Pasterski:2016qvg,Stieberger:2018edy}. The celestial operators associated to these wavefunctions transform as primaries under the action of the conformal group on the sphere,\footnote{It has been observed that these primaries realize an enlarged group of symmetries of the scattering problem \cite{Strominger:2017zoo,Pate:2019mfs,Himwich:2019dug,Himwich:2020rro,Pate:2019lpp,Guevara:2019ypd,Guevara:2021abz,Strominger:2021lvk,Himwich:2021dau,Jiang:2021ovh,Banerjee:2020kaa,Pasterski:2021fjn,Pasterski:2020pdk,Pasterski:2021dqe,Narayanan:2020amh,Banerjee:2020vnt,Banerjee:2020zlg,Nandan:2019jas,Hu:2021lrx,Ferro:2021dub,Pano:2021ewd}.} but have continuous boost weight $\Delta\in 1+i\mathbb{R}$ in contrast with the discrete spectrum expected for standard CFTs \cite{Pasterski:2017kqt}.

Constructed in this way, it is a priori unclear what the spectrum and OPE data of a CCFT should be. In standard CFTs these are often read off from two and three-point correlation functions, which play the role of building blocks. However, in CCFTs the particular nature of these objects makes such connection obscure. Indeed, in order to understand fundamental properties such as the Operator Product Expansion (OPE) \cite{Fan:2019emx,Pate:2019lpp} it has proven fruitful to study higher-multiplicity correlation functions instead, hence ignoring the role of two or three-point functions.

Recent analyses \cite{Law:2020xcf,Atanasov:2021cje,Fan:2021isc,Chang:2021wvv,Lam:2017ofc,Fan:2021pbp,Kulp} have shown that the four-point function in CCFTs contains non-trivial information of the spectrum, and in some cases can be related to the three-point structure constants. In particular, the decomposition of \cite{Atanasov:2021cje} has found that not only conformal primaries with $\Delta\in 1+i\mathbb{R}$ are exchanged in the four-point function, but also their so-called light-ray transforms \cite{Kravchuk:2018htv} with continuous and complex spin (see also \cite{Fan:2021isc,Fan:2021pbp}). These operators also appear to play a key role in a recently found symmetry algebra associated to gravitational and gluon scattering \cite{Guevara:2021abz,Sharma:2021gcz,Strominger:2021lvk,Himwich:2021dau,Jiang:2021ovh}. The essential consideration leading to these findings is the analytic continuation of Minkowski scattering amplitudes to $(2,2)$ signature, the so-called Klein space, for which celestial holography is realized by a conformal theory on a $(1,1)$ torus instead of the Riemann sphere \cite{Atanasov:2021oyu,Crawley:2021ivb}.

The goal of this work is three-fold. First, we aim to elucidate the explicit relation between three-point functions and OPE data in CCFT, focusing on the case of massless-to-massive scalar scattering as studied in \cite{Lam:2017ofc,Atanasov:2021cje,Kulp}. Second, we aim to understand light-ray operators as a direct consequence of the CCFT formulation in terms of three-point functions, rather than as an a posteriori finding. Third, we aim to clarify the relation between the OPE data in $(3,1)$ and $(2,2)$ signatures, by means of adopting a precise prescription for analytic continuation.

The main tool we employ is a Poincare-invariant object we term the \textit{celestial OPE block}. For scalar wavefunctions it realizes the exact operator algebra of two massless primaries converging into a massive one. It can be written as

\begin{equation}
    O_{\Delta_1} (z_1,\bar{z}_1)  O_{\Delta_2} (z_2,\bar{z}_2) = \int d\Omega_3 \mathcal{K}(\Delta_i,z_i,\bar{z}_i)   O^{m}_{\Delta_3 }  (z_3,\bar{z}_3) \,, \label{eq:blck-sq}
\end{equation}
where the superscript $m$ indicates that $O^{m}$ is constructed from a massive wavefunction. The measure $d\Omega_3= d\Delta_3dz_3d\bar{z}_3 (\Delta_3-1)^2$ is related to the normalization of such wavefunctions. The integration runs over $\Delta\in 1+i\mathbb{R}$, corresponding to the weight of the conformal primary wavefunctions, and realizes the fact that they generate the space of scattering solutions (up to zero mode subtleties \cite{Donnay:2018neh} in the massless case). However, it is known that for the massive case such a basis is certainly overcomplete due to the presence of shadow wavefunctions \cite{Pasterski:2017kqt}. We will show that this leads to a redundancy in the operator expression \eqref{eq:blck-sq}, and can be used to restrict the integration to $\Delta\in 1+i\mathbb{R}^+$. Notwithstanding, the former expression has the advantage of setting the primary operators and its shadows on equal footing, suggesting that there is no preferred basis.

Depending on the spacetime signature, the integration of $z,\bar{z}$ can cover different topologies. For scattering in Minkowski space it corresponds to the usual (Euclidean) Riemann sphere. For Klein space, however, it corresponds to a particular Poincare patch of the Lorentzian cylinder as defined in \cite{Kravchuk:2018htv}. We will see how the structure of the cylinder and its patches arise from a simple starting point: the existence of branch cuts in the the three-point functions of CCFT. We will further argue that this is nothing but the universal cover of the celestial torus introduced in \cite{Atanasov:2021oyu} two dimensions. One of the main results of this paper is the relation between the OPE block in both signatures: Motivated by the derivation of the KLT relation between open and closed string integrals \cite{Kawai:1985xq}, we will show how the Euclidean OPE block can be analytically continued to the Lorentzian block. We will then proceed to show how restricting the block to different integration regions leads directly to OPE contributions from the primary operators, their shadows, and their light-ray transforms in the Lorentzian case. These operators are locally independent and hence should all be considered when introducing a local OPE expansion, thereby providing an underlying explanation for the findings of \cite{Atanasov:2021cje}.

The coefficient $\mathcal{K}(\Delta_i,z_i,\bar{z}_i) $ in \eqref{eq:blck-sq} is directly related to the OPE data. Moreover, it can be determined easily in terms of the three-point functions of the theory, or equivalently, it can be completely fixed using Poincare invariance. The latter follows from the fact that for the measure $d\Omega_3$ we find the identity

\begin{equation}
     \int d\Omega_3 \mathcal{K}(\Delta_i,z_i,\bar{z}_i)  P^\mu_{\Delta_3} O^{m}_{\Delta_3 }  (z_3,\bar{z}_3) =  \int d\Omega_3 P^\mu_{2{-}\Delta_3}[\mathcal{K}(\Delta_i,z_i,\bar{z}_i)]   O^{m}_{\Delta_3 }  (z_3,\bar{z}_3) \,,
\end{equation}
where $P^\mu_\Delta$ is the translation generator acting on a conformal primary of weight $\Delta$, as constructed from a massive 4d wavefunction. Then, covariance of the OPE block under translations and Lorentz transformations leads to the same constraints used in \cite{Law:2019glh} to fix the three-point data, here realized by $\mathcal{K}(\Delta_i,z_i,\bar{z}_i)$.

This paper is organized as follows. In section \ref{sec:eucblock} we will construct and explicitly evaluate the Euclidean OPE block in CCFT from three-point data, showing how to account for the contributions of the conformal primaries and their shadow transforms. In section \ref{sec:analyticcon} we study the analytic continuation of the three-point functions to Lorentzian cylinder, also accounting for the closely related role of the 1d light-ray transform. Finally, in section \ref{sec:lorblccft} we apply the analytic continuation to define the Lorentzian OPE block, and extract from it the OPE data together with the light-ray operators. In appendices \ref{app:shadows} and \ref{app:pairings} we discuss essential formulae for light and shadow transforms in CCFT, whereas in appendix \ref{app:distinct} we analyze Poincare covariance of the OPE block for distinct weights.

\section{Euclidean block in CCFT}\label{sec:eucblock}

In the following we will be interested in the contribution of a massive
primary $O_{1+\nu}^{m}(z_{2}), \nu \in i\mathbb{R},$ to the OPE of two massless primaries
$O_{\Delta_1}(z_{1}),O_{\Delta_2}(z_{2})$. Here massless (massive) means
that their correlation functions are obtained from the massless (massive)
wavefunctions of \cite{Pasterski:2017kqt}. Schematically, correlation functions follow from the general formula

\begin{align}
    \langle O_{\Delta_1}(z_{1})O_{\Delta_2}(z_{2})\ldots   O_{1+i\nu}^{m}(z_{i})\ldots \rangle =& \int d\omega_1 \omega^{\Delta_1 -1 } \int d\omega_2 \omega^{\Delta_2 -1 }\ldots  \int \frac{d^3p_i}{p^0_i} \varphi^m_{1+\nu }(z_i;p_i)\ldots \nonumber \\
    & \mathcal{T}(\omega_1 q_1^\mu(z_1),\omega_2 q_2^\mu (z_2),\ldots ,mp_i^\mu,\ldots) \,,
\end{align}
where $q_j^2=0, p_i^2=1$ and $ \mathcal{T}$ is a (distributional, momentum-space) scattering amplitude for massless and massive scalars. See e.g. \cite{Lam:2017ofc} for more details.

It was further observed in \cite{Pasterski:2017kqt} that the massive conformal wavefunctions $\varphi_{\Delta}^{m}(z_i;p_i)$
with $\Delta\in1+i\mathbb{R}$ form
a basis of wavefunctions for the scattering problem.\footnote{We may refer to $\Delta\in1+i\mathbb{R}$ as the principal series. However, it should be clearly distinguished from the \textit{unitary principal series} of $SL(2,\mathbb{C})$ representations. Indeed, the primaries we are considering here lead to highest-weight representations which are non-unitary in the usual sense.} Together with unitarity of the scattering amplitude, this suggests that the OPE contraction of $O_{\Delta_1}(z_{1})$ and $O_{\Delta_2}(z_{2})$
into a massive operator $O_{1+\nu}^{m}(z_{2})$ should involve
the contribution from the full principal series. Let us focus from now on on the case $\Delta_1=\Delta_2$, relegating the case of different weights to the Appendix \ref{app:distinct}. The previous consideration leads to the conjectural form

\begin{equation}
O_{\Delta}(z_{1})O_{\Delta}(z_{2})\sim\int_{-i\infty}^{i\infty}d\nu\frac{F(\nu)}{|z_{12}|^{2\Delta-1-\nu}}O_{1+\nu}^{m}(z_{2})\,, \label{eq:OPEprop}
\end{equation}
which should be understood to hold inside a CCFT correlation function, as we will check explicitly in certain examples.
We will refer to $F(\nu)$ as the OPE data. 

The authors of \cite{Pasterski:2017kqt}
also observed that wavefunction solutions $\varphi_{1+i\nu}^{m}$
with $\nu<0$ are related to those with $\nu>0$ via a shadow transform.
We confirm this in Appendix \ref{app:shadows} by showing that the corresponding
primaries are indeed linearly dependent, with the precise relation
given by 

\begin{equation}
\tilde{O}_{1+\nu}^{m}(z):=\int\frac{d^{2}z_{P}}{2\pi}\frac{O_{1+\nu}^{m}(z_{P})}{|z-z_{P}|^{2-2\nu}}=\frac{1}{\nu}O_{1-\nu}^{m}(z)\,, \label{eq:blsh}
\end{equation}
inside correlation functions. Note that such relation is non-local,
meaning that point-wise $O_{1-\nu}^{m}(z)$ is indeed independent
from $O_{1+\nu}^{m}(z)$ and an OPE of the form (\ref{eq:OPEprop}),
which counts both contributions, makes sense.

Now, in this section we will be interested in the extension of (\ref{eq:OPEprop})
that includes the exact dependence in coordinates $z_{1},z_{2}$ and
not only the leading term. Such object is fixed by global conformal
symmetry, which in the Euclidean setup corresponds to the action
of $SL(2,\mathbb{C})$. This means it includes the contribution from
the complete $SL(2,\mathbb{C})$ descendant family of $O_{1+\nu}^{m}(z_{2})$
and hence usually receives the name of conformal OPE block \cite{Czech:2016xec}.

For $SL(2,\mathbb{C})$ conformal symmetry, the OPE block associated to (\ref{eq:OPEprop})
takes the non-local form:

\begin{equation}
O_{\Delta}(z_{1})O_{\Delta}(z_{2})=\int_{-i\infty}^{i\infty}d\nu\frac{K'(\nu)}{|z_{12}|^{2\Delta}}\times\int\frac{d^{2}z_{P}}{2\pi}\frac{|z_{12}|^{1-\nu}O_{1+\nu}^{m}(z_{P})}{|z_{P1}|^{1-\nu}|z_{P2}|^{1-\nu}}\,,\label{eq:block}
\end{equation}
Note that the integral in $z_{P}$ is conformally invariant for arbitrary
$z_{1},z_{2}$, whereas the full block is conformally covariant as anticipated. The coefficient $K'(\nu)$ of the conformal block
is indeed related to the OPE data $F(\nu)$ apearing in (\ref{eq:OPEprop}),
as explained in the next section. Furthermore, in the notation of equation \eqref{eq:blck-sq} we have

\begin{equation}
    \mathcal{K}(\Delta_i,z_i,\bar{z}_i)=\frac{K'(\nu)}{2\pi \nu^2 }\times\frac{ |z_{12}|^{1-\nu-2\Delta}}{|z_{31}|^{1-\nu}|z_{32}|^{1-\nu}}\,,\quad \Delta_1=\Delta_2=\Delta\,, \Delta_3=1+\nu\,.
\end{equation}
whereas the general form of $ \mathcal{K}(\Delta_i,z_i,\bar{z}_i)$ can be read off from \eqref{eq:genblc}.

The advantage of the block (\ref{eq:block}) over the OPE \eqref{eq:OPEprop} is that the contribution
from the shadow primaries $\tilde{O}_{1+\nu}$ can be easily incorporated.
This is because the block considers the contribution of the primaries
$O_{1+\nu}^{m}(z_{P})$ at all points $z_{P}\in\mathbb{C}^{*}$,
so we can make use of the linear dependence (\ref{eq:blsh}) relating
$\nu\in i\mathbb{R}^{+}$ with $\nu\in i\mathbb{R}^{-}$. It is easy to
convince oneself that this turns (\ref{eq:block}) into 

\begin{equation}
O_{\Delta}(z_{1})O_{\Delta}(z_{2})=\int_{0}^{i\infty}d\nu\frac{K(\nu)}{|z_{12}|^{2\Delta}}\times\int\frac{d^{2}z_{P}}{2\pi}\frac{|z_{12}|^{1-\nu}O_{1+\nu}^{m}(z_{P})}{|z_{P1}|^{1-\nu}|z_{P2}|^{1-\nu}}\,,\label{eq:blockkv1}
\end{equation}
where we have restricted the integration to $\nu\in i\mathbb{R}^{+}$. 

Note that (\ref{eq:blockkv1}) involves a different function than
(\ref{eq:block}), i.e. $K(\nu)\neq K'(\nu)$. Intuitively,
there should be no preferred choice between the cases $\nu\in i\mathbb{R}^{+}$
or $\nu\in i\mathbb{R}^{-}$, meaning that we should be able to write
the complementary relation

\begin{equation}
O_{\Delta}(z_{1})O_{\Delta}(z_{2})=\int_{0}^{i\infty}d\nu\frac{\tilde{K}(\nu)}{|z_{12}|^{2\Delta}}\times\int\frac{d^{2}z_{P}}{2\pi}\frac{|z_{12}|^{1+\nu}O_{1-\nu}^{m}(z_{P})}{|z_{P1}|^{1+\nu}|z_{P2}|^{1+\nu}}\,,\label{eq:kblockkv2}
\end{equation}
where no preferred choice means $\tilde{K}(\nu)=K(-\nu).$ With this
in mind, note further that averaging expressions (\ref{eq:blockkv1})
and (\ref{eq:kblockkv2}) would lead to

\begin{equation}
K(\nu)=2K'(\nu)\,.\label{eq:kv2kv}
\end{equation}
We will confirm this result at the end of this section by showing explicitly
that $\tilde{K}(\nu)=K(-\nu)$.

We aim now to test our expressions for the OPE block and to determine the OPE data. Let us focus first on the form of the block given in (\ref{eq:blockkv1}). The function
$K(\nu)$ is completely fixed by providing the 2-point and 3-point
correlation functions of the theory. In celestial CFT, these are given
by \cite{Lam:2017ofc}

\begin{equation}
\langle O_{\Delta}(z_{1})O_{\Delta}(z_{2})O_{\Delta_{3}}^{m}(z_{3})\rangle=\frac{C(\Delta,\Delta,\Delta_{3})}{|z_{12}|^{2\Delta-\Delta_{3}}|z_{13}|^{\Delta_{3}}|z_{23}|^{\Delta_{3}}}\,,
\end{equation}
with
\begin{equation}
    C(\Delta,\Delta,\Delta_{3})=D_{3}\frac{\Gamma(\Delta_{3}/2)^{2}}{\Gamma(\Delta_{3})}=\frac{gm^{2\Delta-3}}{2^{2\Delta}}\frac{\Gamma(\Delta_{3}/2)^{2}}{\Gamma(\Delta_{3})}\label{eq:3pt}\,.
\end{equation}
The normalization constant $D_3$ depends on the external weights and will be irrelevant for our discussion. Furthermore, we show in Appendix \ref{app:shadows}, using the linear relation (\ref{eq:blsh}),
that the 2-point function is constrained to be \footnote{Importantly, this differs by a factor of 2 in the first term with
respect to the 2-point pairing found in \cite{Pasterski:2017kqt} using bulk
to boundary propagators. Our additional factor of 2 arises from the
distribution $\delta^{2}(z)$ defined as conjugate to the measure
$d^{2}z=2idxdy$ instead of $dxdy$.}

\begin{equation}
\langle O_{\Delta_{1}}^{m}(z_{1})O_{\Delta_{2}}^{m}(z_{2})\rangle=D_{2}\left[-\frac{2\pi\delta(i(\Delta_{1}+\Delta_{2}-2))\delta^{2}(z_{12})}{(\Delta_{1}-1)^{2}}+\frac{1}{(\Delta_{1}-1)}\frac{\delta(i(\Delta_{1}-\Delta_{2}))}{|z_{12}|^{2\Delta_{1}}}\right]\,,\label{eq:2pt}
\end{equation}
for some normalization constant $D_2$. Now let us contract (\ref{eq:blockkv1}) with $O_{\Delta_{3}}^{m}(z_{3})$.
Note that if we choose $\Delta_{3}=1+\nu_{3},\nu_{3}\in i\mathbb{R}^{+}$
only the second term in the above 2-point function will contribute.
We obtain

\begin{align}
\langle O_{\Delta}(z_{1})O_{\Delta}(z_{2})O_{\Delta_{3}}^{m}(z_{3})\rangle & =\int_{0}^{i\infty}d\nu\frac{K(\nu)}{|z_{12}|^{2\Delta}}\int\frac{d^{2}z_{P}}{2\pi}\frac{|z_{12}|^{1-\nu}\langle O_{1+\nu}^{m}(z_{P})O_{\Delta_{3}=1+\nu_{3}}^{m}(z_{3})\rangle}{|z_{P1}|^{1-\nu}|z_{P2}|^{1-\nu}}\nonumber \\
 & =\frac{iD_{2}}{\nu_{3}}\frac{K(\nu_{3})}{|z_{12}|^{2\Delta-1+\nu}}\int\frac{d^{2}z_{P}}{2\pi}\frac{1}{|z_{P1}|^{1-\nu_{3}}|z_{P2}|^{1-\nu_{3}}|z_{P3}|^{2+2\nu_{3}}}\nonumber \\
 & =\frac{iD_{2}}{\nu_{3}}\frac{K(\nu_{3})}{|z_{12}|^{2\Delta-\Delta_{3}}|z_{13}|^{\Delta_{3}}|z_{23}|^{\Delta_{3}}}\times\int\frac{d^{2}z_{P}}{2\pi}\frac{|z_{12}|^{-2\nu_{3}}|z_{13}|^{1+\nu_{3}}|z_{23}|^{1+\nu_{3}}}{|z_{P1}|^{1-\nu_{3}}|z_{P2}|^{1-\nu_{3}}|z_{P3}|^{2+2\nu_{3}}}\nonumber \\
 & =\frac{iD_{2}}{-\nu_{3}^{2}}\frac{K(\nu_{3})}{|z_{12}|^{2\Delta-\Delta_{3}}|z_{13}|^{\Delta_{3}}|z_{23}|^{\Delta_{3}}}\frac{\Gamma(\frac{\Delta_{3}}{2})^{2}\Gamma(2-\Delta_{3})}{\Gamma(\frac{2-\Delta_{3}}{2})^{2}\Gamma(\Delta_{3})}\,.\label{eq:derK}
\end{align}
In the third line we have arranged the integral so that it becomes
conformally invariant, after which we can set i.e. $z_{1}=0,z_{2}=1,z_{3}\to\infty$
which turns it into the familiar integral for a closed string amplitude
\cite{Polchinski:1998rq}. Now the function $K(\nu)$ is fixed by comparing to (\ref{eq:3pt}),
which gives

\begin{equation}
K(\nu)=\frac{iD_{3}}{D_{2}}(\Delta-1)^{2}\frac{\Gamma(\frac{2-\Delta}{2})^{2}}{\Gamma(2-\Delta)}=\frac{iD_{3}}{D_{2}}(\Delta-1)^{2}B_{2-\Delta}\quad,\Delta=1+\nu,\nu\in i\mathbb{R}^{+}.\label{eq:Kvv}
\end{equation}
Here we introduced the shorthand notation $B_{\Delta}=B(\Delta/2,\Delta/2)$. 

Note that the 3-point function (\ref{eq:3pt}) makes no distinction
between $\nu_{3}\in i\mathbb{R}^{+}$ and $\nu_{3}\in i\mathbb{R}^{-}$.
As a consistency check, we proceed to contract the OPE block (\ref{eq:blockkv1})
with $O_{\Delta_{3}}^{m}(z_{3})$, but choosing $\Delta_{3}=1+\nu_{3},\nu_{3}\in i\mathbb{R}^{-}$
this time. Now only the first term in (\ref{eq:2pt}) contributes
instead, and so we obtain

\begin{align}
\langle O_{\Delta}(z_{1})O_{\Delta}(z_{2})O_{\Delta_{3}}^{m}(z_{3})\rangle & =\int_{0}^{i\infty}d\nu\frac{K(\nu)}{|z_{12}|^{2\Delta}}\int\frac{d^{2}z_{P}}{2\pi}\frac{|z_{12}|^{1-\nu}\langle O_{1+\nu}^{m}(z_{P})O_{\Delta_{3}=1+\nu_{3}}^{m}(z_{3})\rangle}{|z_{P1}|^{1-\nu}|z_{P2}|^{1-\nu}}\nonumber \\
 & =\frac{-i2\pi D_{2}}{\nu_{3}{}^{2}}\frac{K(-\nu_{3})}{|z_{12}|^{2\Delta}}\times\frac{1}{2\pi}\frac{|z_{12}|^{1+\nu_{3}}}{|z_{31}|^{1+\nu_{3}}|z_{32}|^{1+\nu_{3}}}\nonumber \\
 & =\frac{D_{3}B_{\Delta}}{|z_{12}|^{2\Delta-1-\nu_{3}}|z_{31}|^{1+\nu_{3}}|z_{32}|^{1+\nu_{3}}}\,,
\end{align}
in perfect agreement with (\ref{eq:3pt}). It is now evident that
the same derivation can be repeated for the `shadow block' (\ref{eq:kblockkv2}).
This simply requires to change $\nu\to-\nu$ in each step of (\ref{eq:derK}),
using the block (\ref{eq:kblockkv2}) instead, leading to the result

\begin{equation}
\tilde{K}(\nu)=\frac{iD_{3}}{D_{2}}(\Delta-1)^{2}B_{\Delta}=K(-\nu)\quad\Delta=1+\nu,\nu\in i\mathbb{R}^{+}\,,
\end{equation}
therefore justifying (\ref{eq:kv2kv}). Note that the function $K(\nu)$,
obtained in (\ref{eq:Kvv}), has been analytically extended to all $\nu\in i\mathbb{R}$.
This is a direct consequence of the CCFT 3-point function (\ref{eq:3pt})
being analytic in this region and meromorphic in the full $\nu$ plane.

\subsection{Singling out contributions in the OPE block}\label{sec:extract}

As anticipated, we can use the singularity structure of the conformal
block to define the local OPE data $F(\nu)$ for the principal series
$\nu\in i\mathbb{R}$. This leads to the form (\ref{eq:OPEprop})
which should hold inside correlation functions of the CCFT, as will be explicitly checked.

To construct the local form of the block we first elaborate on how
precisely the expression (\ref{eq:blockkv1}) can account locally
for the primary $O_{1+\nu}$ and its shadow $\tilde{O}_{1+\nu}$.
This adapts similar discussions for the case of partial waves, see
e.g. \cite{Karateev:2018oml}. We anticipate this construction will be also
useful in the Lorentzian continuation since the block will also contain
light-ray states in the same manner as it contains the shadow states. 

First, we note that the integral

\begin{equation}
\int\frac{d^{2}z_{P}}{2\pi}\frac{|z_{12}|^{1-\nu}O_{1+\nu}^{m}(z_{P})}{|z_{P1}|^{1-\nu}|z_{P2}|^{1-\nu}}\label{eq:intreg}
\end{equation}
receives contributions from a region where 1) $z_{P}$ is close to
either $z_{1},z_{2}$ and 2) $z_{P}$ is far from both $z_{1},z_{2}$.
Region 1) will dominate in the OPE limit involving the primary $O_{1+\nu}^{m}(z_{P})$.
In order to extract such contribution we can ``blow up'' such region
via the change of variables

\begin{equation}
z_{1P}=z_{12}z\,\,z_{P2}=z_{12}(1-z)\label{eq:blowup}\,,
\end{equation}
and then take $z_{12}\to0$ in the integral. This leads to 

\begin{align}
\int\frac{d^{2}z_{P}}{2\pi}\frac{|z_{12}|^{1-\nu}O_{1+\nu}^{m}(z_{P})}{|z_{P1}|^{1-\nu}|z_{P2}|^{1-\nu}}\supset& \nonumber \\ |z_{12}|^{1+\nu}O_{1+\nu}^{m}(z_{2})&\int\frac{d^{2}z}{2\pi}\frac{1}{|z|^{1-\nu}|1-z|^{1-\nu}}=|z_{12}|^{1+\nu}O_{1+\nu}^{m}(z_{2})\frac{\Gamma(\frac{1+\nu}{2})^{2}\Gamma(-\nu)}{\Gamma(\frac{1-\nu}{2})^{2}\Gamma(1+\nu)}\,.\label{eq:singc}
\end{align}
We can now effectively discard this leading term in the OPE limit
by considering region 2) instead. It is easy to see that to integrate
over region 2) we simply take the limit $z_{1}\to z_{2}$ inside the
integral (\ref{eq:intreg}).\footnote{In the case of conformal partial waves, region 2) is equivalently
defined either as $z$ being far from $z_{1},z_{2}$ or probing the
vicinity of the remaining puntures, say $z_{3},z_{4}$ \cite{Karateev:2018oml}.
Formally speaking, the blow-up procedure used to explore these regions
(\ref{eq:blowup}) is an instance of Deligne-Mumford compactification
of punctured Riemann spheres \cite{Deligne1969}.} This immediately gives 

\begin{equation}
\int\frac{d^{2}z_{P}}{2\pi}\frac{|z_{12}|^{1-\nu}O_{1+\nu}^{m}(z_{P})}{|z_{P1}|^{1-\nu}|z_{P2}|^{1-\nu}}\supset|z_{12}|^{1-\nu}\tilde{O}_{1+\nu}^{m}(z_{2})\,.
\end{equation}
We see that different regions can be used to single out the contribution
from the primaries and their shadows. Accounting for both contributions
in the block (\ref{eq:blockkv1}) then leads to the expression

\begin{equation}
O_{\Delta}(z_{1})O_{\Delta}(z_{2})\sim\int_{0}^{i\infty}d\nu\times\left[\frac{F(\nu)}{|z_{12}|^{2\Delta-1-\nu}}O_{1+\nu}^{m}(z_{2})+\frac{K(\nu)}{|z_{12}|^{2\Delta-1+\nu}}\tilde{O}_{1+\nu}^{m}(z_{2})\right]\,,\label{eq:bothcon}
\end{equation}
where 
\begin{equation}
F(\nu)=\frac{\Gamma(\frac{1+\nu}{2})^{2}\Gamma(-\nu)}{\Gamma(\frac{1-\nu}{2})^{2}\Gamma(1+\nu)}K(\nu)=\frac{B_{\Delta}}{B_{2-\Delta}}\frac{K(\nu)}{1-\Delta}\,.\label{eq:FandK}
\end{equation}

From the second term in (\ref{eq:bothcon}) we see that $K(\nu)$ corresponds to the shadow OPE data, rather
than the standard OPE data. The analogous result is well-known for partial waves in standard CFT \cite{Simmons-Duffin:2017nub}. The relation between
the two forms of OPE data is indeed given by (\ref{eq:FandK}) and has been recently
studied in \cite{Pate:2020}. 

Finally, to manifest the symmetry between shadows and their primaries
we can use the explicit form of $K(\nu$). Plugging (\ref{eq:Kvv}),
(\ref{eq:FandK}) into (\ref{eq:bothcon}), using the normalization
(\ref{eq:blsh}) for the shadow, gives
\begin{align}
O_{\Delta}(z_{1})O_{\Delta}(z_{2}) & \sim-\int_{0}^{i\infty}id\nu\times\frac{D_{3}}{D_{2}}\left[\frac{\nu B_{1+\nu}}{|z_{12}|^{2\Delta-1-\nu}}O_{1+\nu}^{m}(z_{2})-\frac{\nu B_{1-\nu}}{|z_{12}|^{2\Delta-1+\nu}}O_{1-\nu}^{m}(z_{2})\right]\nonumber \\
 & =-\frac{D_{3}}{D_{2}}\int_{-i\infty}^{i\infty}i\nu d\nu\frac{B_{1+\nu}O_{1+\nu}^{m}(z_{2})}{|z_{12}|^{2\Delta-1-\nu}}\,.\label{eq:leadn}
\end{align}
This then confirms the structure of the proposed expression (\ref{eq:OPEprop}).
This form of the OPE should be valid inside correlation functions.
For instance, contracting (\ref{eq:leadn}) with $O_{1+\nu_{3}}^{m}(z_{3})$
and discarding the contact term that arises from the 2-point function
we obtain

\begin{equation}
\langle O_{\Delta}(z_{1})O_{\Delta}(z_{2})O_{\Delta_{3}=1+\nu_{3}}^{m}(z_{3})\rangle\sim D_{3}\frac{B_{\Delta_{3}}}{|z_{12}|^{2\Delta-1-\nu}|z_{23}|^{2\Delta_{3}}}\,,
\end{equation}
precisely the OPE behavior of the 3-point function. 

Again, we have found that the OPE data is meromorphic in the $\nu$-plane as induced from the CCFT 3-point function. Singularitites are manifest at negative integer weight $\Delta_3$, in consistency with previous analyses \cite{Pate:2019mfs,Guevara:2021abz}. 

\section{Analytic continuation and Poincare patches}\label{sec:analyticcon}

Our aim now is to extend the previous construction to Lorentzian signature. Motivated by the direct relation to scattering amplitudes this appears to be a natural framework for CCFT \cite{Atanasov:2021oyu,Atanasov:2021cje,Crawley:2021ivb}. We will here proceed
by analytic continuation.

Before continuing the OPE block to Lorentzian
signature we study here the singularity structure
of the analytically continued correlation functions.\footnote{For clarity of the argument we shall consider in this section the analytic extension of the functions to generic weights $\Delta_i$, i.e. not restricted to the principal series $1+i\mathbb{R}$.} In the Euclidean case Wightman functions such as $\langle0|O_{1}O_{2}O_{3}|0\rangle$
and $\langle0|O_{1}O_{3}O_{2}|0\rangle$ are simply fixed by conformal
invariance and hence are equivalent. In the Lorentzian case they can
have a different singularity structure which plays an important role
when one of the operators is a so-called light-ray operator. 

The singularity structure of correlation functions can be understood as a consequence of the Wick rotation
prescription 

\begin{equation}
|z_{1}-z_{2}|^{2}\to z_{12}\bar{z}_{12}+i\epsilon\,\,,z_{12},\bar{z}_{12}\in\mathbb{R}\,,\label{eq:wickr}
\end{equation}
(i.e. now $z,\bar{z}$ are real variables related to the usual Lorentzian
coordinates by $z=x+t,\bar{z}=x-t$). The continuation yields one-dimensional
singularities as opposed to point-like singularities of the Euclidean
case. For instance, for scalars, consider the correlation function

\begin{equation}
\langle O_{\Delta}O_{\Delta}O_{\Delta_{3}}\rangle=\frac{N}{(z_{12}\bar{z}_{12}+i\epsilon)^{\Delta-\Delta_{3}/2}(z_{13}\bar{z}_{13}+i\epsilon)^{\Delta_{3}/2}(z_{23}\bar{z}_{23}+i\epsilon)^{\Delta_{3}/2}}\,.\label{eq:lorcor}
\end{equation}
Note that the correlation function \textsl{does not} factorize into
a left and a right part due to the $i\epsilon$ prescription. This
effectively marks a distinction between correlation functions in Lorentzian
2D to those of 1D+1D, even though they both transform under the conformal
group $SL(2,\mathbb{R})\times SL(2,\mathbb{R})$. The function is
singular when $z_{ij}=0$ or $\bar{z}_{ij}=0$ which corresponds to
two operators being null separated. Consider for instance the case
$\bar{z}_{1}<\bar{z}_{3}<\bar{z}_{2}$, $z_{1}<z_{2}$, which in particular
means that $O_{\Delta}(z_{1}),O_{\Delta}(z_{2})$ are spacelike separated.
Regarding the expression (\ref{eq:lorcor}) as a function of $z_{3}$,
it adopts the form

\begin{equation}
\langle O_{\Delta}O_{\Delta}O_{\Delta_{3}}\rangle=\frac{N}{(z_{12}\bar{z}_{12})^{\Delta-\Delta_{3}/2}\bar{z}_{13}^{\Delta_{3}/2}\bar{z}_{23}^{\Delta_{3}/2}(z_{13}-i\epsilon)^{\Delta_{3}/2}(z_{23}+i\epsilon)^{\Delta_{3}/2}}\,,\label{eq:3ptlasfun}
\end{equation}
which has branch points located at $z_{3}=z_{1}-i\epsilon,z_{3}=z_{2}+i\epsilon$.
The monodromies of the branch points are both equal to $e^{i\pi\Delta_{3}}$,
this means that there exists another branch point at infinity with
monodromy $e^{-2i\pi\Delta_{3}}$. For reasons that will
become obvious in the next section branch cuts should not ``overlap''
and neither cross the horizontal axis; hence we can only
assign them as follows: The left cut corresponds to the interval $(-\infty,z_{1})$
and lies slightly below the horizontal axis, while the right cut corresponds
to the interval $(z_{2},+\infty)$ and lies above such axis. The discontinuities
across such branch cuts are then related to the commutators $[O_{\Delta}(z_{1}),O_{\Delta_{3}}(z_{3})]$
and $[O_{\Delta}(z_2),O_{\Delta_{3}}(z_{3})]$ respectively, and indeed
vanish when $z_{3}>z_{1}$ or $z_{3}<z_{2}$ (respectively) since $O_{\Delta_{3}}(z_{3})$
becomes timelike separated from them. We provide an explicit example
of this by considering light-ray operators in the next subsection.

\begin{figure}[h]
\centering
\includegraphics[width=5cm]{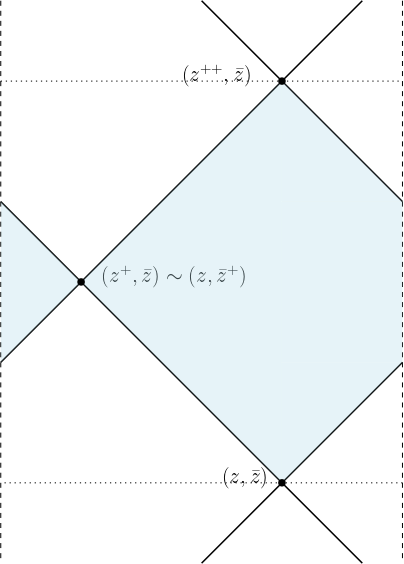}
\caption{\textit{The Lorentzian cylinder and a colored Poincare patch in the real $z,\bar{z}$ slice. Dashed lines are identified, see \cite{Kravchuk:2018htv} for details on the construction. Further identifying the dotted lines leads to the Lorentzian torus.}}\label{fig:cylx}
\end{figure}

The existence of the branch cuts means that correlation functions for generic weights
are defined on a multi-sheeted Riemann covering our patch
$\mathbb{R^{*}\times\mathbb{R}^{*}}$. Indeed, the patch $\mathbb{R^{*}\times\mathbb{R}^{*}}$ for real $z,\bar{z}$ is an instance of the
the Poincare patches studied in \cite{Kravchuk:2018htv}. To construct the multi-sheeted surface, consider a generic 2D Lorentzian correlation and follow
it analytically from $z_{1}\to-\infty$ to $z_{1}\to\infty$, that
is, along a null geodesic. The function then has a discontinuity reflecting
the change of Poincare patch. Indeed, due to conformal symmetry it
is clear that

\begin{equation}
\langle O_{\Delta}\ldots\rangle\to\frac{C}{(\pm|z_{1}|)^{\Delta}}\,\,\,,\,\,z_{1}\to\pm\infty\,.\label{eq:infm}
\end{equation}
Hence in general there is a monodromy factor of $e^{-2\pi i\Delta}$
associated to $|z_{1}|\to\infty$.\footnote{More precisely, to measure the monodromy we proceed as follows: In
going from $z_{1}\to-\infty$ to $z\to+\infty$ according to (\ref{eq:infm})
we pick up a phase $e^{-i\pi\Delta}$. We then return from $+\infty$
to $-\infty$ by following the one-parameter curve $z_{1}=Re^{i\theta}$
(for large radius $R$) and set $\theta\to\pi$, which leads to an
extra $e^{-i\pi\Delta}$.} This is precisely the monodromy factor we associated to $|z_{3}|=\infty$
in our example (\ref{eq:3ptlasfun}), in which case it follows from
the fact that we cross the branch cuts in going from $z_{3}=-\infty$
to $z_{3}=+\infty$. For a general correlation function there will
be as many branch cuts as operators that can become null separated
with $O_{\Delta}$, all of them with an endpoint at $\infty$. The choice of
Poincare patch in the $z$-plane corresponds to the single complex sheet containing $[-\infty,\infty[$ in this conformal frame, which we refer covariantly as interval $[z_3,z_3^+[$.
The universal cover of such patches is in fact the Lorentzian cylinder
depicted in Figure \ref{fig:cylx}.

So far we have learned that conformal symmetry implies that our generic-weight primaries (more precisely, the irreducible representations) live in the Lorentzian cylinder. To close this section let us elucidate the relation between the cylinder and the celestial torus in $\mathbb{R}^{2,2}$ space, constructed recently in \cite{Atanasov:2021oyu}. For this it is convenient to recast \eqref{eq:infm} in the notation of \cite{Kravchuk:2018htv},

\begin{equation}
    TO_\Delta (z_1) |0\rangle = e^{i\pi \Delta} O_\Delta(z_1) |0\rangle\,,
\end{equation}
where $T$ is an operator generating discrete null translations along the cylinder, such that $T[z_1]=z_1^+$ for constant $\bar{z}$. Now, ref. \cite{Atanasov:2021oyu} constructed highest-weight representations associated to the conformal primaries of \textit{integer} weight $\Delta$. Acting on such particular primaries we observe that $T^2 =\mathbb{I}$ and hence we identify $z_1\sim z_1^{++}$ in the cylinder: This restores time periodicity and quotients the cylinder into the torus, see Figure \ref{fig:cylx}.  

\subsection{Light-ray transform}\label{sec:LRT}

Light-ray operators are intimately tied to the above discussion and correspond to an important
avatar of Lorentzian CFTs. We define the light transform for
a scalar operator as

\begin{equation}
L[O_{\Delta}](z,\bar{z})=\int_{z}^{z^{+}}\frac{dz_{P}}{(z_{P}-z)^{2-\Delta}}O_{\Delta}(z_{P},\bar{z})\,.\label{eq:ltre}
\end{equation}
From covariance of the integrand we find that $L[O_{\Delta}]$ has
weight $\Delta_{L}=1$ and continuous spin $J_{L}=1-\Delta$. In \cite{Kravchuk:2018htv}
such a transformation has been thoroughly studied in general dimensions.
Here we will adapt part of that treatment to two dimensions and rephrase
it to make connection with the sheeted Riemann surfaces we have just
identified.

First, we take $z,z^{+}$ in (\ref{eq:ltre}) to define a Poincare
patch, which means that the contour is precisely the null geodesic
we considered in the previous section, going from $z_{P}=z$ to $z_{P}=z^{+}$
for fixed $\bar{z}$. Notice that in this case however, the branch
point defining the patch is not located at infinity but precisely
at $z_{P}=z$, although we can choose a conformal frame such that
$z=-\infty,z^{+}=+\infty$. The fact that there is no $i\epsilon$
prescription for this branch point at infinity reflects the fact that
the contour formally does not cross $z_{P}=z,z^{+}$.

\begin{figure}[h]
\centering
\includegraphics[width=10cm]{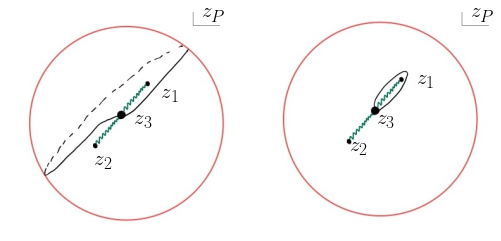}
\caption{\textit{Left: The integration contour over a projective
real line on the Riemann sphere, starting and ending at $z_{3},z_{3}^{+}$
respectively. Right: Deformation of the contour towards the left
cut. Note that the contour does not intersect any branch cut nor branch
point.}}\label{fig:sph}
\end{figure}

In order to understand the singularity structure of these objects
let us briefly study the 3-point functions $\langle O_{\Delta}O_{\Delta}L[O_{\Delta3}]\rangle$.
The time ordered correlation function is discontinuous and defined
piecewise. For definiteness let us consider the operators $O_{\Delta}(z_{1}),O_{\Delta}(z_{2})$
to be spacelike separated, i.e. $\bar{z}_{1}<\bar{z}_{2},z_{1}<z_{2}$.
We also anticipate that $z_{3}$ will define a certain Poincare patch
so we will further assume $z_{3}<z_{1}<z_{2}$. In such case we have
only three domains for the time ordered function:

\begin{equation}
\langle O_{\Delta}O_{\Delta}L[O_{\Delta_{3}}]\rangle=\begin{cases}
\langle0|O_{\Delta}(z_{1})O_{\Delta}(z_{2})L[O_{\Delta_{3}}]|0\rangle & \bar{z}_{1}<\bar{z}_{2}<\bar{z}_{3}\\
\langle0|O_{\Delta}(z_{1})L[O_{\Delta_{3}}]O_{\Delta}(z_{2})|0\rangle & \bar{z}_{1}<\bar{z}_{3}<\bar{z}_{2}\\
\langle0|L[O_{\Delta_{3}}]O_{\Delta}(z_{1})O_{\Delta}(z_{2})|0\rangle & \bar{z}_{3}<\bar{z}_{1}<\bar{z}_{2}
\end{cases}\label{eq:regions}
\end{equation}
Each of the objects on the right-hand side is indeed an analytic function defined as follows:
It is computed in its corresponding domain, using the $i\epsilon$
prescription (\ref{eq:wickr}), and then analytically extended to
the other domains. For instance

\begin{align}
\langle0|O_{\Delta}(z_{1})O_{\Delta}(z_{2})L[O_{\Delta_{3}}]|0\rangle & =\frac{N}{(z_{12}\bar{z}_{12})^{\Delta-\Delta_{3}/2}\bar{z}_{13}^{\Delta_{3}/2}\bar{z}_{23}^{\Delta_{3}/2}}\int_{z_{3}}^{z_{3}^{+}}\frac{dz_{P}}{z_{P3}{}^{2-\Delta_{3}}}\frac{1}{(z_{1P}-i\epsilon)^{\Delta_{3}/2}(z_{2P}-i\epsilon)^{\Delta_{3}/2}}\nonumber \\
\langle0|O_{\Delta}(z_{1})L[O_{\Delta_{3}}]O_{\Delta}(z_{2})|0\rangle & =\frac{N}{(z_{12}\bar{z}_{12})^{\Delta-\Delta_{3}/2}\bar{z}_{13}^{\Delta_{3}/2}\bar{z}_{23}^{\Delta_{3}/2}}\int_{z_{3}}^{z_{3}^{+}}\frac{dz_{P}}{z_{P3}{}^{2-\Delta_{3}}}\frac{1}{(z_{1P}-i\epsilon)^{\Delta_{3}/2}(z_{2P}+i\epsilon)^{\Delta_{3}/2}}\label{eq:2ints}
\end{align}
Note that the latter is precisely the function studied in (\ref{eq:3ptlasfun})
integrated along the Poincare patch we defined, once we set the frame
$z_{3}=-\infty,z_{3}^{+}=+\infty$. The corresponding contour is depicted
in the Riemman sphere in Figure \ref{fig:sph} (left). Now, from (\ref{eq:2ints})
we can see that in the time ordered function $\langle O_{\Delta}O_{\Delta}L[O_{\Delta3}]\rangle$
passing analytically from the first region in (\ref{eq:regions}) to
the second region causes the branch cut starting at $z_{P}=z_{2}$
to ``pinch'' the integration contour. More precisely the discontinuity
occurs when $\bar{z}_{3}\to\bar{z}_{2}$ (i.e. $O_{\Delta}(z_{2}),L[O_{\Delta3}]$
become null separated) and is given by the difference between the
above analytic functions: 

\begin{equation}
\langle0|O_{\Delta}(z_{1})[O_{\Delta}(z_{2}),L[O_{\Delta_{3}}]]|0\rangle=\frac{N}{(z_{12}\bar{z}_{12})^{\Delta-\Delta_{3}/2}\bar{z}_{13}^{\Delta_{3}/2}\bar{z}_{23}^{\Delta_{3}/2}}\int_{(z_{2},z_{3}^{+})}\frac{dz_{P}}{z_{P3}{}^{2-\Delta_{3}}z_{1P}{}^{\Delta_{3}/2}z_{2P}{}^{\Delta_{3}/2}}\,,\label{eq:com3}
\end{equation}
where the notation $(z_{2},z_{3}^{+})$ in the integral means we must
hug the right branch cut. 

It is easy to see that the above integrals in $z_{P}$ are conformally
covariant (or invariant with appropriate normalization), which is
equivalent to the statement that the monodromies of the integrand
$e^{2i\pi(2-\Delta_{3})},e^{i\pi\Delta_{3}},e^{i\pi\Delta_{3}}$ multiply
to the identity. A consequence of this is that there are no other
singularities besides the ones that are shown explicitly and the contour
can be deformed freely. For instance, in the first integral the contour
lies above both branch cuts and can be closed from above. So we find
i.e.

\begin{equation}
\langle0|O_{\Delta}(z_{1})O_{\Delta}(z_{2})L[O_{\Delta_{3}}]|0\rangle=\langle0|L[O_{\Delta_{3}}]O_{\Delta}(z_{1})O_{\Delta}(z_{2})|0\rangle=0\,,
\end{equation}
which is the statement that the light-ray operators annihilate the
vacuum \cite{Kravchuk:2018htv}. In particular this means that the time ordered
correlator (\ref{eq:regions}) is only non-trivial when $\bar{z}_{1}<\bar{z}_{3}<\bar{z}_{2}$,
in which case it is given by the second line of (\ref{eq:2ints}).
In such integral we can close the contour from below or above as depicted
in Figure \ref{fig:sph}, leading to the equivalent expressions

\begin{equation}
\langle O_{\Delta}O_{\Delta}L[O_{\Delta_{3}}]\rangle=\langle0|O_{\Delta}(z_{1})[O_{\Delta}(z_{2}),L[O_{\Delta_{3}}]]|0\rangle=\langle0|[O_{\Delta}(z_{1}),L[O_{\Delta3}]]O_{\Delta}(z_{2})|0\rangle\,,
\end{equation}

It will be useful for the following to compute this explicitly. Let
us look at the second expression, which corresponds to the integral
(\ref{eq:com3}). We can write it as
\begin{align}
\langle O_{\Delta}O_{\Delta}L[O_{\Delta_{3}}]\rangle & {=}\frac{N}{\bar{z}_{21}{}^{\Delta-\Delta_{3}/2}\bar{z}_{31}^{\Delta_{3}/2}\bar{z}_{23}^{\Delta_{3}/2}z_{21}^{\Delta-1+\Delta_{3}/2}z_{23}^{1-\Delta_{3}/2}z_{13}^{1-\Delta_{3}/2}}\int_{(z_{2},z_{3}^{+})}dz_{p}\frac{z_{21}^{\Delta_{3}{-}1}z_{23}^{1{-}\Delta_{3}/2}z_{13}^{1{-}\Delta_{3}/2}}{z_{p3}{}^{2-\Delta_{3}}z_{p1}{}^{\Delta_{3}/2}z_{2p}{}^{\Delta_{3}/2}}\nonumber \\
 & {=}\frac{N \times 2i\sin\pi\Delta_{3}/2 }{\bar{z}_{21}{}^{\Delta-\Delta_{3}/2}\bar{z}_{31}^{\Delta_{3}/2}\bar{z}_{23}^{\Delta_{3}/2}z_{21}^{\Delta-1+\Delta_{3}/2}z_{23}^{1-\Delta_{3}/2}z_{13}^{1-\Delta_{3}/2}}\int_{1}^{\infty}dt\frac{1}{t^{\Delta_{3}/2}(t-1){}^{\Delta_{3}/2}}\nonumber \\
 & {=}\frac{N}{\bar{z}_{21}{}^{\Delta-\Delta_{3}/2}\bar{z}_{31}^{\Delta_{3}/2}\bar{z}_{23}^{\Delta_{3}/2}z_{21}^{\Delta-1+\Delta_{3}/2}z_{23}^{1-\Delta_{3}/2}z_{13}^{1-\Delta_{3}/2}}\times2\pi i\frac{\Gamma(\Delta_{3}-1)}{\Gamma(\Delta_{3}/2)^{2}}\,.\label{eq:00l}
\end{align}
In the first line we wrote the integration in a manifestly conformally
invariant form, after which we are free to set $z_{3}=-\infty,z_{1}=0,z_{2}=1$
inside the integral. This is consistent with our assumption that
$z_{3}<z_{1}<z_{2}$ which enables us to naturally consider the Poincare
patch starting at $z_{3}$. The factor of $\sin\pi\Delta_{3}/2$ in
the second line comes from writing $z_{2p}=e^{\pm i\pi}(t-1)$ for
above and below the branch cut respectively. The result can be written
as

\begin{equation}
\langle O_{\Delta}O_{\Delta}L[O_{\Delta_{3}}]\rangle=\frac{2\pi i}{(\Delta_{3}-1)B_{\Delta_{3}}}\langle O_{\Delta}O_{\Delta}O_{\Delta_{L}=1,J_{L}=1-\Delta_{3}}\rangle\,,\label{eq:3ptlight}
\end{equation}
where the correlation function in the RHS involves the operator $O_{\Delta_{L},J_{L}}$
with the same OPE data of $O_{\Delta_{3}}$, given by the normalization
$N$ above. Thus, this is essentially a relation between the OPE data
of $O_{\Delta_{3}}$ and its light-ray transform, as we will see explicitly
using the OPE block. In the case of CCFT, with $N=D_{3}B_{\Delta_{3}}$
as given by (\ref{eq:3pt}), we obtain

\begin{equation}
\langle O_{\Delta}O_{\Delta}L[O_{\Delta_{3}}^{m}]\rangle=\frac{2\pi i}{(\Delta_{3}-1)}\frac{D_{3}}{\bar{z}_{21}{}^{\Delta-\Delta_{3}/2}\bar{z}_{31}^{\Delta_{3}/2}\bar{z}_{23}^{\Delta_{3}/2}z_{21}^{\Delta-1+\Delta_{3}/2}z_{23}^{1-\Delta_{3}/2}z_{13}^{1-\Delta_{3}/2}}\,.\label{eq:3ptlightf}
\end{equation}
We will see how this result is consistent with the OPE block in the next sections.

\section{Lorentzian block in CCFT}\label{sec:lorblccft}

We are now in position to study the Lorentzian OPE conformal block
in CCFT. We will construct this object by analytic continuation of
the Euclidean case following the prescription in the last section.
We start from our Euclidean expression (\ref{eq:blockkv1}) and perform
the usual Wick rotation. Using $d^{2}z_{P}\to-idz_{P}d\bar{z}_{P}$
we obtain

\begin{equation}
O_{\Delta}(z_{1})O_{\Delta}(z_{2})=\int_{0}^{i\infty}\frac{d\nu}{2\pi i}\frac{K(\nu)}{z_{12}^{\Delta}\bar{z}_{12}^{\Delta}}\int_{\mathbb{R}}\frac{d\bar{z}_{p}\bar{z}_{21}^{1-h}}{\bar{z}_{p1}^{1-h}\bar{z}_{p2}^{1-h}}\int_{\mathbb{R}}\frac{dz_{p}z_{21}^{1-h}}{z_{p1}^{1-h}z_{p2}^{1-h}}O_{2h}^{m}(z_{p},\bar{z}_{p})\qquad h=\frac{1+\nu}{2}\label{eq:lbck}
\end{equation}
with the $i\epsilon$ prescription implicit. Recall this means that
the integral does not factorize into left and right blocks, even though
each of the components transform covariantly under $SL(2,\mathbb{R})$.
To actually compute this expression as two integrals over the real
line we need to choose a Poincare patch for which both contours are
defined. We will see that we can fix the patch at will, as long as
the choice is consistent between the two integrals.

We now want to test the analytic continuation by contracting the block
against a massive primary, which should lead us to the (analytically
continued) 3-point function (\ref{eq:3pt}). For this, we contract
(\ref{eq:lbck}) with $O_{\Delta_{3}}^{m}(z_{p},\bar{z}_{P})$, $\Delta_{3}=1+\nu_{3},\nu_{3}\in i\mathbb{R}^{+}$and
again use the second term in the 2-point function (\ref{eq:2pt}).
The result is

\begin{align}
\langle O_{\Delta}(z_{1})O_{\Delta}(z_{2})O_{\Delta_{3}}^{m}(z_{3})\rangle & =\frac{1}{2\pi}\frac{K(\nu_{3})}{z_{12}^{\Delta}\bar{z}_{12}^{\Delta}}\frac{iD_{2}}{\nu_{3}}\times\frac{\bar{z}_{21}^{h}z_{21}^{h}}{\bar{z}_{32}^{h}\bar{z}_{31}^{h}z_{32}^{h}z_{31}^{h}}\label{eq:bfr}\\
 & \qquad\int_{\mathbb{R}}\frac{d\bar{z}_{p}\bar{z}_{12}^{1-2h}\bar{z}_{32}^{h}\bar{z}_{31}^{h}}{\bar{z}_{p1}^{1-h}\bar{z}_{p2}^{1-h}\bar{z}_{3p}^{2h}}\int_{\mathbb{R}}\frac{dz_{p}z_{12}^{1-h}z_{32}^{h}z_{31}^{h}}{z_{p1}^{1-h}z_{p2}^{1-h}z_{3p}^{2h}}\qquad h=\frac{1+\nu_{3}}{2}\,.
\end{align}
In order to compute the integrals we define the endpoints
of the contours. Let us for instance take the patch to be defined
by setting its origin in $z_{3},\bar{z}_{3}$. It is then convenient to assume without loss
of generality that $z_{3}<z_{1}<z_{2},\bar{z}_{3}<\bar{z}_{1}<\bar{z}_{2}$
so that all of the punctures lie in the same patch. The integrals
then take the form

\begin{equation}
\int_{\bar{z}_{3}}^{\bar{z}_{3}^{+}}\frac{d\bar{z}_{p}\bar{z}_{21}^{1-2h}\bar{z}_{23}^{h}\bar{z}_{13}^{h}}{\bar{z}_{p1}^{1-h}\bar{z}_{p2}^{1-h}\bar{z}_{p3}^{2h}}\int_{z_{3}}^{z_{3}^{+}}\frac{dz_{p}z_{21}^{1-h}z_{23}^{h}z_{13}^{h}}{z_{p1}^{1-h}z_{p2}^{1-h}z_{p3}^{2h}}\,,
\end{equation}
where we are free to drop the $i\epsilon$ prescription at $z_{p}=z_{3},\bar{z}_{p}=\bar{z}_{3}$
singularities. The strategy is now to split the $\bar{z}_{3}$ integration
in three regions:

\begin{equation}
\bar{z}_{p}<\bar{z}_{1}\,,\,\,\bar{z}_{1}<\bar{z}_{p}<\bar{z}_{2}\,\,\,\textrm{and}\,\,\bar{z}_{2}<\bar{z}_{p}\,.
\end{equation}
For each of these regions the other integral takes precisely one of
the three forms we discussed in the previous section. More precisely,
due to the $i\epsilon$ prescription, we easily find that when $\bar{z}_{p}<\bar{z}_{1}$
the contour on the right integral can be closed from above, and when
$\bar{z}_{2}<\bar{z}_{p}$ it can be closed from below, giving zero
in both cases. It is only in the region $\bar{z}_{1}<\bar{z}_{p}<\bar{z}_{2}$
that we obtain the non-trivial contribution

\begin{equation}
\int_{\bar{z}_{1}}^{\bar{z}_{2}}\frac{d\bar{z}_{p}\bar{z}_{21}^{1-2h}\bar{z}_{23}^{h}\bar{z}_{13}^{h}}{\bar{z}_{p1}^{1-h}\bar{z}_{2p}^{1-h}\bar{z}_{p3}^{2h}}\times\int_{z_{3}}^{z_{3}^{+}}\frac{dz_{p}z_{21}^{1-h}z_{23}^{h}z_{13}^{h}}{(z_{p1}+i\epsilon)^{1-h}(z_{2p}+i\epsilon)^{1-h}z_{p3}^{2h}}\,.
\end{equation}
Now the integrals take a factorized form and we can compute them separately:
The integral on the right-hand side was computed in the previous section
by closing the contour to either of the cuts. The left integral takes
the defining form of the beta function by using the gauge fixing $\bar{z}_{3}=-\infty,\bar{z}_{1}=0,\bar{z}_{2}=1$
consistent with $\bar{z}_{3}<\bar{z}_{1}<\bar{z}_{2}$. The result
of the above expression is then

\begin{align}
\int_{\mathbb{R}}\frac{d\bar{z}_{p}\bar{z}_{12}^{1-2h}\bar{z}_{32}^{h}\bar{z}_{31}^{h}}{\bar{z}_{p1}^{1-h}\bar{z}_{p2}^{1-h}\bar{z}_{3p}^{2h}}\int_{\mathbb{R}}\frac{dz_{p}z_{12}^{1-h}z_{32}^{h}z_{31}^{h}}{z_{p1}^{1-h}z_{p2}^{1-h}z_{3p}^{2h}} & =\int_{\bar{z}_{1}}^{\bar{z}_{2}}\frac{d\bar{z}_{p}\bar{z}_{21}^{1-2h}\bar{z}_{23}^{h}\bar{z}_{13}^{h}}{\bar{z}_{p1}^{1-h}\bar{z}_{2p}^{1-h}\bar{z}_{p3}^{2h}}\times2i\sin\pi h\int_{z_{2}}^{z_{3}^{+}}\frac{dz_{p}z_{21}^{1-h}z_{23}^{h}z_{13}^{h}}{z_{p1}^{1-h}z_{p2}^{1-h}z_{p3}^{2h}}\nonumber\\
 & =2\pi i\frac{\Gamma(h)^{2}}{\Gamma(2h)}\frac{\Gamma(1-2h)}{\Gamma(1-h)^{2}}\nonumber \\
 & =\frac{2\pi i}{1-2h}\times\frac{B_{2h}}{B_{2-2h}}\,. \label{eq:ancon}
\end{align}
Plugging this back into (\ref{eq:bfr}), using (\ref{eq:Kvv}), we
obtain

\begin{equation}
\langle O_{\Delta}(z_{1})O_{\Delta}(z_{2})O_{\Delta_{3}}^{m}(z_{3})\rangle=\frac{D_{3}B_{2h}}{z_{12}^{\Delta}\bar{z}_{12}^{\Delta}}\frac{\bar{z}_{21}^{h}z_{21}^{h}}{\bar{z}_{32}^{h}\bar{z}_{31}^{h}z_{32}^{h}z_{31}^{h}}\,\,,\,h=\frac{\Delta_{3}}{2}\,,
\end{equation}
as desired.

The outcome of (\ref{eq:ancon}) was expected. In fact, comparing
this to the analogous derivation done in the Euclidean case, eq. (\ref{eq:derK}),
shows that our result is equivalent to the statement

\begin{align}
\int d^{2}z_{P}\frac{|z_{12}|^{-2\nu_{3}}|z_{13}|^{1+\nu_{3}}|z_{23}|^{1+\nu_{3}}}{|z_{P1}|^{1-\nu_{3}}|z_{P2}|^{1-\nu_{3}}|z_{P3}|^{2+2\nu_{3}}} & =\int_{\bar{z}_{1}}^{\bar{z}_{2}}\frac{d\bar{z}_{p}\bar{z}_{21}^{1-2h}\bar{z}_{23}^{h}\bar{z}_{13}^{h}}{\bar{z}_{p1}^{1-h}\bar{z}_{2p}^{1-h}\bar{z}_{p3}^{2h}}\times2\sin\pi h\int_{z_{2}}^{z_{3}^{+}}\frac{dz_{p}z_{21}^{1-h}z_{23}^{h}z_{13}^{h}}{z_{p1}^{1-h}z_{p2}^{1-h}z_{p3}^{2h}}\,.\\
\nonumber 
\end{align}
which is nothing but the $n=4$ KLT formula familiar from string amplitudes.
Our derivation is closely related to the original analytic continuation given
by KLT in \cite{Kawai:1985xq}, but our additional
discussion about Poincare patches makes the procedure more transparent for our purposes.
In particular in \cite{Kawai:1985xq} the Poincare patch
is implicitly chosen by sending some of the punctures to infinity,
which drops $i\epsilon$ regulator as a byproduct. We have instead
proceeded in a manifestly covariant way, which should enable us to
extend this procedure to more general correlation functions. Furthermore,
the analysis of Poincare patches plays a key role in identifying the
contributions from the light-ray operators in the OPE block, which
we do next.

\subsection{Singling out the light-ray operators}

In the same manner as we have contracted the OPE block with a scalar
primary, one can perform the contraction with a light-ray operator
to obtain the correlation function $\langle O_{\Delta}(z_{1})O_{\Delta}(z_{2})L[O_{\Delta_{3}}](z_{3})\rangle$
given in (\ref{eq:3ptlightf}). For this it is necessary to first
study the 2-point function $\langle L[O_{\Delta}^{m}]O_{\Delta}^{m}\rangle$.
In Appendix \ref{app:pairings} we study such pairings and we show that indeed the analytically
continued OPE block (\ref{eq:lbck}) recovers the correlation function
involving light-ray operators. This in an indirect proof that the
block contains such operators. Here we will provide a direct argument
instead, following the lines of Section \ref{sec:extract}.

As explained in Section \ref{sec:extract} the integration of the conformal block
can be split into two regions. In the first region $z_{P}(\bar{z}_{P})$
probes the vicinity of $z_{1},z_{2}$ (or $\bar{z}_{1},\bar{z}_{2})$,
whereas in the second region $z_{P}$ is integrated far from both
punctures. To single out the contribution from the light-ray operator
we will consider the first region in the $\bar{z}_{P}$ integral,
and the second region in the $z_{P}$ integral. The second region
can be singled out easily from taking the coincident limit $z_{12}\to0$
inside integration, thus (\ref{eq:lbck}) takes the following form (assume $z_2>z_1,\bar{z}_2>\bar{z}_1$ as in section \ref{sec:LRT}).

\begin{align}
O_{\Delta}(z_{1})O_{\Delta}(z_{2}) & \supset\int_{0}^{i\infty}\frac{d\nu}{2\pi i}\frac{K(\nu)}{z_{21}^{\Delta+h-1}\bar{z}_{21}^{\Delta}}\int_{\mathbb{R}}\frac{d\bar{z}_{p}\bar{z}_{21}^{1-h}}{\bar{z}_{p1}^{1-h}\bar{z}_{p2}^{1-h}}\int_{\mathbb{R}}\frac{dz_{p}}{z_{p2}^{2-2h}}O_{2h}^{m}(z_{p},\bar{z}_{P})\nonumber \\
 & =\int_{0}^{i\infty}\frac{d\nu}{2\pi i}\frac{K(\nu)}{z_{21}^{\Delta+h-1}\bar{z}_{21}^{\Delta}}\int_{\mathbb{R}}\frac{d\bar{z}_{p}\bar{z}_{21}^{1-h}}{\bar{z}_{p1}^{1-h}\bar{z}_{p2}^{1-h}}L[O_{2h}^{m}](z_{2},\bar{z}_{P})\,,\label{eq:lightap}
\end{align}
Notice that we have implicitly taken the Poincare patch to start at
$z_{P}=z_{1}=z_{2}$. We now want to extract the local contribution
from the remaining integral, obtained from $z_{P}$ being in the vicinity
of $z_{1},z_{2}$. Instead of defining the blow-up as in section \ref{sec:extract},
let us proceed in a slightly different manner. To explain it, we take a step back and revisit the Euclidean case. A direct way of
extracting the local contribution $z_{P}\sim z_{1},z_{2}$ is simply
to replace $O_{1+\nu}^{m}(z_{P})=O_{1+\nu}^{m}(z_{2})+\ldots$ in
the Euclidean block, where the corrections given in $\ldots$ can
be regarded as descendants an hence ignored. Thus 

\begin{align}
\int\frac{d^{2}z_{P}}{2\pi}\frac{|z_{12}|^{1-\nu}O_{1+\nu}^{m}(z_{P})}{|z_{P1}|^{1-\nu}|z_{P2}|^{1-\nu}} & \supset O_{1+\nu}^{m}(z_{2})\int\frac{d^{2}z_{P}}{2\pi}\frac{|z_{12}|^{1-\nu}}{|z_{P1}|^{1-\nu}|z_{P2}|^{1-\nu}}\label{eq:exfd}\\
 & =O_{1+\nu}^{m}(z_{2})|z_{12}|^{1+\nu}\int\frac{d^{2}z_{P}}{2\pi}\lim_{z_{3}\to\infty}\frac{|z_{12}|^{-2\nu}|z_{13}|^{1+\nu}|z_{23}|^{1+\nu}}{|z_{P1}|^{1-\nu}|z_{P2}|^{1-\nu}|z_{P3}|^{2+2\nu}}\\
 & =O_{1+\nu}^{m}(z_{2})|z_{12}|^{1+\nu}\frac{\Gamma(\frac{1+\nu}{2})^{2}\Gamma(-\nu)}{\Gamma(\frac{1-\nu}{2})^{2}\Gamma(1+\nu)}\,.
\end{align}
This gives the same result as the blow-up argument (\ref{eq:singc}), but the argument is more transparent for our purposes: The integral was computed
when we evaluated the 3-point function in (\ref{eq:derK}). Note that the coefficient of $O_{1+\nu}^{m}(z_{2})$ in (\ref{eq:exfd})
can be understood as the 3-point function $z_{3}^{2\Delta_{3}}\langle O_{\Delta}(z_{1})O_{\Delta}(z_{2})O_{1+\nu}(z_{3})\rangle$
with $z_{1}\to z_{2}$ and $z_{3}\to\infty$, which is precisely how
the coefficient of a certain primary appearing in the $O_{\Delta}(z_{1})O_{\Delta}(z_{2})$
OPE can be extracted. In adapting this reasoning to the Lorentzian
case, we learn that to extract the singular piece we can include an
extra puncture $z_{3}$ in the block for which we will then take $|z_{3}|\to\infty$,
meaning that it will correspond to the origin of our Poincare patch. 

Thus, let us now consider the previous reasoning applied to the $\bar{z}_{P}$
integration in (\ref{eq:lightap})

\begin{align}
\int_{\mathbb{R}}\frac{d\bar{z}_{p}\bar{z}_{21}^{1-h}}{\bar{z}_{p1}^{1-h}\bar{z}_{p2}^{1-h}}L[O_{2h}^{m}](z_{2},\bar{z}_{P}) & \supset L[O_{2h}^{m}](z_{2},\bar{z}_{2})\int_{\mathbb{R}}\frac{d\bar{z}_{p}\bar{z}_{21}^{1-h}}{\bar{z}_{p1}^{1-h}\bar{z}_{p2}^{1-h}}\nonumber \\
 & =L[O_{2h}^{m}](z_{2},\bar{z}_{2})\bar{z}_{21}^{h}e^{i\pi (h-1)}\lim_{\bar{z}_{3}\to-\infty}\int_{\bar{z}_{3}}^{\bar{z}_{3}^{+}}\frac{d\bar{z}_{p}\bar{z}_{21}^{1-2h}\bar{z}_{13}^{h}\bar{z}_{23}^{h}}{\bar{z}_{p1}^{1-h}\bar{z}_{2p}^{1-h}\bar{z}_{p3}^{2h}}\nonumber \\
 & =L[O_{2h}^{m}](z_{2},\bar{z}_{2})\bar{z}_{21}^{h}e^{i\pi (h-1)}\times2\pi i\frac{\Gamma(1-2h)}{\Gamma(1-h)^{2}}\,,\label{eq:sndf}
   \end{align}
where we have made explicit that the integral must be done in the
Poincare patch of $\bar{z}_{3}$. Again, the integral has been computed
in (\ref{eq:00l}) for the case of the 3-point function $\langle OOL\rangle$,
with the appropriate $i\epsilon$ prescriptions. Indeed, the `phase' factor $e^{i\pi (h-1)}$, or more generally $e^{i\pi\frac{2h+\Delta_1-\Delta_2-2}{2}}$, has been identified in \cite{Kravchuk:2018htv} as associated to such three-point correlator and will be important to recover equation \eqref{eq:3ptlightf} as we will now show.

Plugging this back
into (\ref{eq:lightap}), using (\ref{eq:Kvv}), gives

\begin{equation}
O_{\Delta}(z_{1})O_{\Delta}(z_{2})\supset\frac{iD_{3}}{D_{2}}\int_{0}^{i\infty}d\nu\frac{1-2h}{z_{21}^{\Delta+h-1}\bar{z}_{21}^{\Delta-h}}e^{i\pi(h-1)}L[O_{2h}^{m}](z_{2})\,\,\,,h=\frac{1+\nu}{2}\,.\label{eq:oplg}
\end{equation}
This is the main result of this section. The expansion is consistent with the light-ray OPE data obtained in \cite{Atanasov:2021oyu} from a direct decomposition of the four-point amplitude, see discussion, but includes the extra phase factor which is required for the three-point function.

The contour in \eqref{eq:oplg} may be extended to the full imaginary axis by accounting
for the shadow contribution $L[O_{2-2h}^{m}]$, similarly to section
\ref{sec:extract}. We will not carry out such analysis here. Instead, let us just
consider the contraction of the above OPE with another light-ray operator.
The needed two point function is computed in Appendix \ref{app:pairings}, and is given
by eq. (\ref{eq:llc}). For the case $z_2>z_1,\bar{z}_2>\bar{z}_1$ it leads to\footnote{We could also include the singled out contribution from scalars to
the OPE $O_{\Delta}(z_{1})O_{\Delta}(z_{2})$. However, this will
not contribute to this particular contraction, as it follows from
\ref{eq:LOp} that it vanishes when $z_{23}\neq0$. } 

\begin{align}
\langle O_{\Delta}(z_{1})O_{\Delta}(z_{2})L[O_{\Delta_{3}}^{m}](z_{3})\rangle & \sim\frac{iD_{3}}{D_{2}}\int_{0}^{i\infty}d\nu\frac{1-2h}{z_{21}^{\Delta}\bar{z}_{21}^{\Delta-h}}e^{i\pi(h-1)}\langle L[O_{2h}^{m}](z_{2})L[O_{\Delta_{3}}^{m}](z_{3})\rangle\nonumber \\
 & =-\frac{2\pi iD_{3}}{(\Delta_{3}-1)}\frac{e^{i\pi(h-1)}}{z_{21}^{\Delta+\Delta_{3}/2-1}\bar{z}_{21}^{\Delta-\Delta_{3}/2}\bar{z}_{23}^{\Delta_{3}}z_{23}^{2-\Delta_{3}}}\,,h=\Delta_3/2\,.
\end{align}
To see how this matches the leading behavior of the 3-point function (\ref{eq:3ptlightf}) we use that $\bar{z}_{31}^{\Delta_3/2}\to e^{-i\pi h}\bar{z}_{23}^{\Delta_3/2}$ as $\bar{z}_1\to\bar{z}_2$. This explains the appearance of the phase factor in the numerator.

\section{Discussion}

In this work we have constructed the celestial conformal block associated to the OPE $O_{\Delta}(z_{1})O_{\Delta}(z_{2})$, for both Euclidean and Lorentzian
signature as related by analytic continuation.
As anticipated, the analysis can be easily extended to the case of distinct weights,
with the treatment of the analytic continuation being essentially
identical. We provide the explicit form of this celestial OPE block in Appendix \ref{app:distinct}.

To understand unitarity and constructibility in CCFTs it is important to establish the relation between higher-point correlation functions and the 3-pt functions studied here. In \cite{Atanasov:2021cje} a conformal block decomposition of the 4-pt function
was performed in Lorentzian signature. It was found that the t-channel
expansion involves massive scalar primaries with dimensions $\Delta=n+1\,\,,n\geq1$
and light-ray operators of dimension $\Delta=1$ and spin $J\in i\mathbb{R}$.
The light-ray operators extracted from our
OPE block (\ref{eq:oplg}) carry the same coefficients as the ones
found in that reference, up an overall normalization and a phase factor $e^{i\pi(h-1)}$. The latter should cancel in the pairing between the light-ray operator and its shadow to build the 4-pt function. It would be further interesting to recover the partial wave expression of \cite{Atanasov:2021cje} from our OPE block. As the former exhibits an explicit factorization in terms of $sl(2,\mathbb{R})_L\times sl(2,\mathbb{R})_R$ partial waves this would further require to factorize the contracted OPE blocks similarly to what was done here at three-points.

On the other hand, here we have relied on the conformal primary operators with $\Delta\in 1+i\mathbb{R}$ to write down the expression for the OPE block. Even though this leads to a complete set of operators organized in a Poincare invariant form, it is still not clear what the precise basis should be. Accounting for the shadow transforms and the light-ray transforms would certainly lead to an overcomplete basis. More importantly, their associated highest-weight representations are well-known to be non-unitary for such range of $\Delta$, see the recent discussion in \cite{Liu:2021tif}. This suggests that even though we are able to write the precise OPE in terms of operators over the principal series, there should exist an analytic continuation in $\Delta$ leading to discrete unitary representations, perhaps valued on the celestial torus. The recent findings of \cite{Atanasov:2021cje} and \cite{Kulp} confirm this at four-points for scalar scattering. For a precise matching to the OPE data it is further required to obtain an analytic continuation of two-point pairings such as \eqref{eq:2pt} to the full $\Delta$ plane, see \cite{Donnay:2020guq} for the massless case.

The OPE block for massless-to-massive primaries should be essentially different than its analog for all-massless primaries. In the latter case the OPE coefficients correspond to
a distribution in $\Delta$ and the three-point functions are singular in $z$. Indeed, as we anticipated in the introduction, the OPE block for (positive-helicity) gluons,

\begin{equation}
    O^{+,a}_{\Delta_1}O^{+,b}_{\Delta_2}\supset -if^{abc}\frac{\Gamma(1-\Delta_1)\Gamma(1-\Delta_2)}{2\pi\Gamma(1-\Delta_1-\Delta_2)}\int \frac{d^2 z_3 O^{+,c}_{\Delta_1+\Delta_2-1}(z_3)}{z_{12}^{\Delta_1+\Delta_2}z_{32}^{1-\Delta_1}z_{31}^{1-\Delta_2}\bar{z}_{12}^{\Delta_1+\Delta_2-3}\bar{z}_{32}^{2-\Delta_1}\bar{z}_{31}^{2-\Delta_2}}\,,\label{eq:glblc}
\end{equation}
introduced recently in \cite{Guevara:2021abz}, was indeed motivated\footnote{More precisely, the chiral factor of the block, termed $sl(2,\mathbb{R})_L$ block, was extracted from higher-point amplitudes using BCFW and then extended to include for $sl(2,\mathbb{R)}_R$ descendants. This does not exclude the possibility of additional primaries appearing in the RHS of \eqref{eq:glblc}.} from higher-point amplitudes instead of three-point scattering as done here. It is then a pressing question to understand the relation of this block with the three-point functions $\langle O^{+,a}(\bar{z}_1)O^{+,b}(\bar{z}_2) O^{-,d}(\bar{z}_3)  \rangle$, if any. A priori, this is complicated by the fact that the latter functions, as obtained in e.g. \cite{Pasterski:2017ylz}, are only defined in the region $\bar{z}_{23}\bar{z}_{31}\geq 0$, hence the OPE limit $\bar{z}_{12}\to 0$ is singular. Let us momentarily insist, however, on the relation \eqref{eq:glblc} as providing an analytic continuation of the OPE in the complex $z,\bar{z}$ planes. In Lorentzian signature, following the procedure of the main text, we can easily extract the contribution from the gluon primary and its light-ray transform from \eqref{eq:glblc}. For $z_2>z_1,\bar{z}_2>\bar{z}_1$ we get 
\begin{equation}
      O^{+,a}_{\Delta_1}O^{+,b}_{\Delta_2}  \supset \frac{f^{abc}}{\bar{z}_{21}^{\Delta_{1}+\Delta_{2}-3}z_{21}}L[O_{\Delta_{1}+\Delta_{2}-1}^{+,c}](z_{2,}\bar{z}_{2})e^{i\pi(\Delta_{1}-1)}-iB(\Delta_1-1,\Delta_2-1)\frac{f^{abc}}{z_{21}} O_{\Delta_{1}+\Delta_{2}-1}^{+,c}(z_{2,}\bar{z}_{2})\,, \label{eq:tent}
\end{equation}
We observe that the light-ray term does not involve the beta function corresponding to the OPE data of gluons (this is the OPE analog of \eqref{eq:3ptlight}), precisely as occurs in the three-point function \cite{Law:2019glh,Sharma:2021gcz}. Indeed, conformal symmetry fixes the gluon light-ray pairing

\begin{equation}
    \langle L[O_{\Delta_{1}}^{+c}](z_1,\bar{z}_1)O_{\Delta_{2}}^{-d}(z_2,\bar{z}_2)\rangle=C \times \delta^{cd} \delta(i(\Delta_{1}+\Delta_{2}-2))\delta(z_{12})\frac{1}{\bar{z}_{21}^{3-\Delta_{1}}}\,,
\end{equation}
from which we obtain the colinear limit (for $\bar{z}_2\neq\bar{z}_3$)
\begin{equation}
     \langle O^{+,a}_{\Delta_1}O^{+,b}_{\Delta_2} O^{-,c}_{\Delta_3} \rangle \sim   \delta\left(i\left(\sum_i\Delta_i - 3\right)\right)\frac{ -C/z_{21} f^{abc} \delta(z_{23})}{\bar{z}_{21}^{-\Delta_3}\bar{z}_{32}^{2-\Delta_2}\bar{z}_{23}^{2-\Delta_1}}
\end{equation}
which agrees with the analytic continuation of the three-point function \cite{Law:2019glh,Sharma:2021gcz} if we regulate $-C/z_{21}\to 2\pi \delta(z_{12})$.
This suggests the correlation function indeed contains the exchange of a light-ray operator rather than a gluon primary. It would be interesting to provide a precise treatment of this case, which could bridge the pressing gap between 3-point functions and operator algebras in gauge and gravity theories \cite{Sharma:2021gcz,Himwich:2021dau}.

\subsection*{Acknowledgements}

We thank  E. Himwich, S. Pasterski, M. Pate, S.H. Shao and A. Strominger for insightful discussions, as well as W. Melton and A.M. Raclariu for collaboration on related projects. This work was supported by DOE grant de-sc/000787 and the Black Hole
Initiative at Harvard University, which is funded by grants from the John Templeton Foundation
and the Gordon and Betty Moore Foundation. The author also receives support from the Harvard Society
of Fellows.

\appendix

\section{Relation between primaries and their shadow transforms}\label{app:shadows}

Here we provide an alternative argument for the linear relation between
primaries and their shadows, given by (\ref{eq:blsh}), complementary
to the derivation given in \cite{Pasterski:2017kqt} using wavefunctions.
Let us simply assume the CCFT correlation functions are constrained
solely in terms of 3-point OPE data, and hence we will only check
the validity of the relation in 3-point functions. Consider then the
following correlation function involving a shadow operator

\begin{align}
\langle O_{\Delta}O_{\Delta}\tilde{O}_{1+\nu}^{m}(z_{3})\rangle & =\int\frac{d^{2}z_{P}}{2\pi}\frac{\langle O_{\Delta}O_{\Delta}O_{1+\nu}^{m}(z_{P})\rangle}{|z_{3}-z_{P}|^{2-2\nu}}\,.
\end{align}

We now use the defining form of the 3-point function for CCFT, given
in (\ref{eq:3pt})

\begin{equation}
\langle O_{\Delta}O_{\Delta}O_{\Delta_{3}=1+\nu_{3}}^{m}(z_{P})\rangle=\frac{D_{3}B_{\Delta_{3}}}{|z_{12}|^{2\Delta-\Delta_{3}}|z_{13}|^{\Delta_{3}}|z_{23}|^{\Delta_{3}}}\,,
\end{equation}
with $B_{\Delta_{3}}=B(\Delta_{3}/2,\Delta_{3}/2)$, to obtain

\begin{align}
\langle O_{\Delta}O_{\Delta}\tilde{O}_{\Delta_{3}=1+\nu}^{m}(z_{3})\rangle & =\frac{D_{3}B_{\Delta_{3}}}{|z_{12}|^{2\Delta-\Delta_{3}}}\int\frac{d^{2}z_{P}}{2\pi}\frac{1}{|z_{3P}|^{2-2\nu}|z_{1P}|^{\Delta_{3}}|z_{23}|^{\Delta_{3}}}\nonumber \\
 & =\frac{D_{3}B_{\Delta_{3}}}{|z_{12}|^{2\Delta-\Delta_{3}}}\int\frac{d^{2}z_{P}}{2\pi}\frac{1}{|z_{3P}|^{2-2\nu}|z_{1P}|^{\Delta_{3}}|z_{23}|^{\Delta_{3}}}\nonumber \\
 & =\frac{D_{3}B_{\Delta_{3}}}{|z_{12}|^{2\Delta-\Delta_{3}}}\times\frac{1}{|z_{12}|^{2\nu_{3}}|z_{13}|^{1-\nu_{3}}|z_{23}|^{1-\nu_{3}}}\times\frac{1}{\nu}\frac{B_{2-\Delta_{3}}}{B_{\Delta_{3}}}\nonumber \\
 & =\frac{1}{\nu}\langle O_{\Delta}O_{\Delta}O_{\Delta_{3}=1-\nu_{3}}^{m}(z_{P})\rangle\,.
\end{align}
The integral has been done using the standard methods described in
the main text (c.f. eq (\ref{eq:derK})). The same procedure can be
repeated straightforwardly for the case of different weights, using
the form of $\langle O_{\Delta_{1}}O_{\Delta_{2}}O_{\Delta_{3}=1+\nu_{3}}^{m}(z_{P})\rangle$
given in \cite{Lam:2017ofc}. Thus we have found that

\begin{equation}
\tilde{O}_{1+\nu}^{m}(z):=\int\frac{d^{2}z_{P}}{2\pi}\frac{O_{1+\nu}^{m}(z_{P})}{|z-z_{P}|^{2-2\nu}}=\frac{1}{\nu}O_{1-\nu}^{m}(z)\,,\label{eq:linrl}
\end{equation}

As $n=2,3$ correlation functions are potentially the only required building blocks in CCFT (see e.g. \cite{Guevara:2019ypd} for the massless case), one would like to check this relation at $n=2$ also.
Indeed, we can enforce it and find it leads to a particular pairing
between primaries and shadows. For this we start from the orthogonal
pairing for operators of same dimension, i.e.

\begin{equation}
\langle O_{\Delta_{1}}^{m}O_{\Delta_{2}}^{m}\rangle=\frac{D_{2}\delta(i\Delta_{1}-i\Delta_{2})}{(\Delta_{1}-1)|z_{12}|^{2\Delta_{1}}}\,\,\,,\,\,\Delta_{1},\Delta_{2}\in1+i\mathbb{R}^{+}\,.\label{eq:2res}
\end{equation}
The pole at $\Delta_{1}$ is a signature of CCFT arising from the
momentum space 2-point function having a $\sim 1/\omega$ behavior. $D_{2}$
is a normalization constant that can be computed explicitly but will not be required here. From
the definition of shadow transform in (\ref{eq:linrl}),

\begin{equation}
\langle\tilde{O}_{\Delta_{1}=1+\nu}^{m}O_{\Delta_{2}}^{m}\rangle:=\frac{D_{2}\delta(i\Delta_{1}-i\Delta_{2})}{\Delta_{2}-1}\int\frac{d^{2}z_{P}}{2\pi}\frac{1}{|z_{1P}|^{2-2\nu}|z_{2P}|^{2+2\nu}}\,.
\end{equation}
This 2-point integral can be found in e.g. \cite{Dolan:2011dv}, but in order
to draw the analogy with the Lorentzian case (treated in Appendix
\ref{app:pairings}), let us compute it explicitly. Note that it is conformally covariant
as the weights add up to $4$. However to compute it explicitly we
can use the known form of the 3-point integrals if we introduce an
extra puncture at infinity with weight $\epsilon\to0$ and unfix the
gauge $z_{3}\to\infty$:

\begin{equation}
\int\frac{d^{2}z_{P}}{2\pi}\frac{1}{|z_{1P}|^{2-2\nu}|z_{2P}|^{2+2\nu}}=\frac{1}{|z_{12}|^{2+\epsilon}}\int\frac{d^{2}z_{P}}{2\pi}\frac{|z_{12}|^{2+\epsilon}|z_{13}|^{-2\nu-\epsilon}|z_{23}|^{2\nu}}{|z_{1P}|^{2-2\nu}|z_{2P}|^{2+2\nu+\epsilon}|z_{3P}|^{-\epsilon}}\,,
\end{equation}
which leads us again to a standard closed string integral, giving

\begin{align}
\int\frac{d^{2}z_{P}}{2\pi}\frac{1}{|z_{1P}|^{2-2\nu}|z_{2P}|^{2+2\nu}} & =\lim_{\epsilon\to0}\frac{1}{|z_{12}|^{2+\epsilon}}\times\frac{\Gamma(\nu)\Gamma(-\nu-\epsilon/2)\Gamma(1+\epsilon/2)}{\Gamma(1-\nu)\Gamma(1+\nu+\epsilon/2)\Gamma(-\epsilon/2)}\nonumber \\
 & =\lim_{\epsilon\to0}\frac{-\epsilon/2}{|z_{12}|^{2+\epsilon}}\times\frac{1}{-\nu^{2}}\nonumber \\
 & =\frac{2\pi\delta^{2}(z_{12})}{-\nu^{2}}\,,
\end{align}
where we have used a representation of the delta function.\footnote{It follows from the standard formula $\partial_{\bar{z}}1/z=2\pi\delta^{2}(z)$
after writing the LHS as $\partial_{\bar{z}}(\bar{z}^{-\epsilon}z^{-1-\epsilon})$.} Thus we obtain

\begin{equation}
\langle\tilde{O}_{\Delta_{1}=1+\nu}^{m}O_{\Delta_{2}}^{m}\rangle:=-\frac{2\pi D_{2}\delta(i\Delta_{1}-i\Delta_{2})}{(\Delta_{2}-1)^{3}}\delta^{2}(z_{12})\,.
\end{equation}
Thus, enforcing the linear relation (\ref{eq:linrl}) gives

\begin{equation}
\langle O_{\Delta_{1}=1-\nu_{2}}^{m}O_{\Delta_{2}=1+\nu_{2}}^{m}\rangle:=-\frac{2\pi D_{2}\delta(i(-\nu_{1}+\nu_{2}))}{(\Delta_{2}-1)^{2}}\delta^{2}(z_{12})\,\,\,,\,\,\nu_{1},\nu_{2}\in i\mathbb{R}^{+}\,,
\end{equation}
which extends (\ref{eq:2res}) to the case $\Delta_{1}\in1+i\mathbb{R}^{-}$.
Putting both contributions together we arrive at (\ref{eq:2pt}).

\section{Distinct Weights and Poincare Symmetry}\label{app:distinct}

As mentioned in the main text, our construction can be directly extended
to the case of different conformal weights. In doing so one encounters
essentially the same conformal integrals over the complex sphere and
projective line. We will not repeat such analysis here, instead let
us simply present the corresponding result for the Euclidean block
(\ref{eq:block}) extended to different weights:

\begin{equation}
O_{\Delta_{1}}(z_{1})O_{\Delta_{2}}(z_{2})=\int_{-i\infty}^{i\infty}d\nu\frac{K'_{\Delta_{1},\Delta_{2}}(\nu)}{|z_{12}|^{\Delta_{1}+\Delta_{2}}}\times\int\frac{d^{2}z_{P}}{2\pi}\frac{|z_{12}|^{1-\nu}O_{1+\nu}^{m}(z_{P})}{|z_{P1}|^{1-\nu+\Delta_{12}}|z_{P2}|^{1-\nu-\Delta_{12}}}\,,\label{eq:genblc}
\end{equation}
where 

\begin{equation}
K'_{\Delta_{1},\Delta_{2}}(\nu)=\frac{iD_{3}}{2D_{2}}(\Delta-1)^{2}B\left(\frac{\Delta_{12}+2-\Delta}{2},\frac{\Delta_{21}+2-\Delta}{2}\right)\quad,\Delta=1+\nu,\Delta_{12}=\Delta_{1}-\Delta_{2}\,.
\end{equation}
By identifying the OPE data \cite{Lam:2017ofc}

\begin{equation}
\langle O_{\Delta_{1}}(z_{1})O_{\Delta_{2}}(z_{2})O_{\Delta}^{m}(z_{3})\rangle=D_{3}\frac{B\left(\frac{\Delta_{12}+\Delta}{2},\frac{\Delta_{21}+\Delta}{2}\right)}{|z_{12}|^{\Delta_{1}+\Delta_{2}-\Delta_{3}}|z_{13}|^{\Delta_{1}+\Delta_{3}-\Delta_{2}}|z_{23}|^{\Delta_{2}+\Delta_{3}-\Delta_{1}}}\,,
\end{equation}
we can compactly write the generic block (\ref{eq:genblc}) in the
form

\begin{equation}
O_{\Delta_{1}}(z_{1})O_{\Delta_{2}}(z_{2})=\frac{i}{2D_{2}}\int\frac{d\Delta d^{2}z_{P}}{2\pi}(\Delta-1)^{2}\langle O_{\Delta_{1}}(z_{1})O_{\Delta_{2}}(z_{2})O_{2-\Delta}^{m}(z_{P})\rangle O_{\Delta}^{m}(z_{P})\,.\label{eq:blckgen2}
\end{equation}
The factor of $(\Delta-1)^{2}$ arises due to the normalization of the operators $O_{\Delta}^{m}$.
The result has the form of a $\textrm{SL}(2,\mathbb{C})$ block (in
the so-called shadow representation) integrated along the axis $\textrm{Re(\ensuremath{\Delta)=1}}$.
In this form we can readily check covariance of our OPE block under
the full Poincare group, or conversely, use the symmetry to find the integral kernel $\langle O_{\Delta_{1}}(z_{1})O_{\Delta_{2}}(z_{2})O_{2-\Delta}^{m}(z_{P})\rangle$.

First note that that Lorentz covariance is
manisfest as both sides transform manisfestly with the corresponding
weights in $z_{1},z_{2}$, whereas the weight in $z_{P}$ cancels
in the integration. To check covariance under translations we use the following form of
the generators acting on conformal primaries \cite{Law:2019glh}

\begin{align}
P^{\mu}[O_{\Delta}(z)] & =q^{\mu}(z)e^{\partial_{\Delta}}[O_{\Delta}(z)]\nonumber \\
P^{\mu}[O_{\Delta}^{m}(z)] & =\frac{m}{2}\left(\left(\partial\bar{\partial}q^{\mu}+\frac{\partial q^{\mu}\bar{\partial}+\bar{\partial}q^{\mu}\partial}{\Delta-1}+\frac{q^{\mu}\partial\bar{\partial}}{(\Delta-1)^{2}}\right)e^{-\partial_{\Delta}}+\frac{\Delta q^{\mu}}{\Delta-1}e^{\partial_{\Delta}}\right)[O_{\Delta}^{m}(z)]\,,\label{eq:pom}
\end{align}
where $q^{\mu}(z,\bar{z})$ is the standard null vector pointing at
a point $z,\bar{z}$ in the celestial sphere. Covariance under translations
of the block (\ref{eq:blckgen2}) then reads

\begin{align}
P_{1}^{\mu}O_{\Delta_{1}}(z_{1})O_{\Delta_{2}}(z_{2})+O_{\Delta_{1}}(z_{1})P_{2}^{\mu}O_{\Delta_{2}}(z_{2})=&\nonumber \\
\frac{i}{2D_{2}}\int\frac{d\Delta d^{2}z_{P}}{2\pi}&(\Delta-1)^{2}\langle O_{\Delta_{1}}(z_{1})O_{\Delta_{2}}(z_{2})O_{2-\Delta}^{m}(z_{P})\rangle P_{P}^{\mu}[O_{\Delta}^{m}(z_{P})]\,.\label{eq:pop1}
\end{align}
By using again (\ref{eq:blckgen2}) we can write the LHS of this equation
as

\begin{align}
P_{1}^{\mu}O_{\Delta_{1}}(z_{1})O_{\Delta_{2}}(z_{2})+O_{\Delta_{1}}(z_{1})P_{2}^{\mu}O_{\Delta_{2}}(z_{2}) & \nonumber\\
=\frac{i}{2D_{2}}\int\frac{d\Delta d^{2}z_{P}}{2\pi}(\Delta-1)^{2}&(P_{1}^{\mu}+P_{2}^{\mu})\langle O_{\Delta_{1}}(z_{1})O_{\Delta_{2}}(z_{2})O_{2-\Delta}^{m}(z_{P})\rangle O_{\Delta}^{m}(z_{P})\nonumber \\
  =\frac{i}{2D_{2}}\int\frac{d\Delta d^{2}z_{P}}{2\pi}(\Delta-1)^{2}&\langle O_{\Delta_{1}}(z_{1})O_{\Delta_{2}}(z_{2})P_{P}^{\mu}[O_{2-\Delta}^{m}(z_{P})]\rangle O_{\Delta}^{m}(z_{P})\,, \label{eq:pop2}
\end{align}
where we have used that the three-point function is invariant under
translations for two incoming and one outgoing wavefunctions in momentum
space, see \cite{Law:2019glh},

\begin{equation}
(P_{1}^{\mu}+P_{2}^{\mu}-P_{3}^{\mu})\langle O_{\Delta_{1}}(z_{1})O_{\Delta_{2}}(z_{2})O_{\Delta}^{m}(z_{3})\rangle=0\,.\label{eq:3pttrs}
\end{equation}
The equality between (\ref{eq:pop1}) and (\ref{eq:pop2}) is then
equivalent to the hermiticity of the translation operators in the
sense that

\begin{equation}
\int d\Delta d^{2}z_{P}(\Delta-1)^{2}\times P_{P}^{\mu}[O_{2-\Delta}^{m}(z_{P})]O_{\Delta}^{m}(z_{P})=\int d\Delta d^{2}z_{P}(\Delta-1)^{2}\times O_{2-\Delta}^{m}(z_{P})P_{P}^{\mu}[O_{\Delta}^{m}(z_{P})]\,.\label{eq:hermic}
\end{equation}

Before we provide the proof of the hermiticity condition, let us emphasize that read backwards (i.e. using \eqref{eq:hermic}, \eqref{eq:pop1} and the first line of \eqref{eq:pop2}), the argument shows that Poincare symmetry implies that the integral kernel in \eqref{eq:blckgen2} is constrained by \eqref{eq:3pttrs} up to a function that vanishes under the integral. Such functions shall be disregarded physically since they would imply that the OPE coefficients in \eqref{eq:blckgen2} are not unique, whereas direct comparison with three-point functions shows that they are. Finally, we note that \eqref{eq:3pttrs} together with Lorentz covariance (as implied from \eqref{eq:blckgen2}) completely fixes the coefficients $\langle O_{\Delta_{1}}(z_{1})O_{\Delta_{2}}(z_{2})O_{2-\Delta}^{m}(z_{P})\rangle$ as shown in \cite{Law:2019glh} under similar assumptions.

We close this appendix by providing a proof of \eqref{eq:hermic}. Using the form of (\ref{eq:pom}) acting of $O_{2-\Delta}^{m}(z)$
and integrating by parts in $z,\bar{z}$ we obtain, after some algebra, 

\begin{align}
\int d\Delta d^{2}z_{P}(\Delta-1)^{2}\times P_{P}^{\mu}[O_{2-\Delta}^{m}(z_{P})]O_{\Delta}^{m}(z_{P}) & \nonumber \\
=\frac{m}{2}\int d\Delta d^{2}z_{P}(\Delta-1)^{2}O_{\Delta}^{m}(z_{P})&\left(\left(\partial\bar{\partial}q^{\mu}-\frac{\partial q^{\mu}\bar{\partial}+\bar{\partial}q^{\mu}\partial}{\Delta-1}+\frac{q^{\mu}\partial\bar{\partial}}{(\Delta-1)^{2}}\right)e^{\partial_{\Delta}} \right. \nonumber\\
\left.+\frac{(2-\Delta)q^{\mu}}{1-\Delta}e^{-\partial_{\Delta}}\right)[O_{2-\Delta}^{m}(z_{P})]\nonumber \\
 =\frac{m}{2}\times\int d\Delta d^{2}z_{P}e^{\partial_{\Delta}}[O_{2-\Delta}^{m}(z)]&\left(\partial\bar{\partial}q^{\mu}+\frac{\partial q^{\mu}\bar{\partial}+\bar{\partial}q^{\mu}\partial}{\Delta}+\frac{q^{\mu}\partial\bar{\partial}}{\Delta^{2}}\right)\Delta^{2}O_{\Delta}^{m}(z_{P})\nonumber \\
 +\frac{m}{2}\int d\Delta d^{2}z_{P}e^{-\partial_{\Delta}}[O_{2-\Delta}^{m}(z)]&(\Delta-2)(\Delta-1)q^{\mu}O_{\Delta}^{m}(z_{P})\,,
\end{align}
Note that the pole at $\Delta=1$ introduced by $P_{P}^{\mu}$ cancels
against the measure, which leads to a well defined contour at $\textrm{Re(\ensuremath{\Delta)=1}}.$
We now regard the operators $e^{\pm\partial_{\Delta}}$ as power series
in $(\partial_{\Delta})^{n}$, which can then be integrated by parts
$n$ times. Assuming that correlation functions are damped as $\Delta \to 1\pm i\infty$ \cite{Law:2019glh} so that there is no boundary term, we can then
write the previous expression as

\begin{align}
\int d\Delta d^{2}z_{P}(\Delta-1)^{2}\times P_{P}^{\mu}&[O_{2-\Delta}^{m}(z_{P})]O_{\Delta}^{m}(z_{P})\nonumber\\
 =\frac{m}{2}\int d\Delta d^{2}z_{P}&O_{2-\Delta}^{m}(z)e^{-\partial_{\Delta}}
\left[\left(\partial\bar{\partial}q^{\mu}+\frac{\partial q^{\mu}\bar{\partial}+\bar{\partial}q^{\mu}\partial}{\Delta}+\frac{q^{\mu}\partial\bar{\partial}}{\Delta^{2}}\right)\Delta^{2}O_{\Delta}^{m}(z_{P})\right]\nonumber \\
  +\frac{m}{2}\int & d\Delta d^{2}z_{P}O_{2-\Delta}^{m}(z)q^{\mu}e^{\partial_{\Delta}}\left[(\Delta{-}2)(\Delta{-}1)O_{\Delta}^{m}(z_{P})\right]\nonumber \\
  =\frac{m}{2}\int d\Delta d^{2}z_{P}&(\Delta{-}1)^{2}O_{2{-}\Delta}^{m}(z_{P})\left(\left(\partial\bar{\partial}q^{\mu}{+}\frac{\partial q^{\mu}\bar{\partial}{+}\bar{\partial}q^{\mu}\partial}{\Delta-1}+\frac{q^{\mu}\partial\bar{\partial}}{(\Delta-1)^{2}}\right)e^{-\partial_{\Delta}}\right.\nonumber\\
  \left.+\frac{\Delta q^{\mu}}{\Delta-1}e^{\partial_{\Delta}}\right)&[O_{\Delta}^{m}(z_{P})]\,,
\end{align}
which is precisely (\ref{eq:hermic}). We then conclude that the OPE
block is Poincare covariant. An important caveat however is the following.
We have integrated by parts the shift operators $e^{\partial_{\Delta}}$,
which are defined as the power series

\begin{equation}
e^{\epsilon\partial_{\Delta}}O_{\Delta}^{m}=\sum_{n=0}^{\infty}\frac{\epsilon^{n}}{n!}\partial_{\Delta}^{n}O_{\Delta}^{m}=O_{\Delta+\epsilon}^{m}\,.
\end{equation}
evaluated at $\epsilon=1$. Of course, the series requires the existence
of a convergence radius $r>1$, which is not true if correlation functions
involving the operator $O_{\Delta+\epsilon}^{m}$ have singularities
for some $0\leq\epsilon\leq1$ at fixed $\Delta=1+i\lambda$. For
instance, depending of the UV behaviour of the theory we may encounter
singularities at $\epsilon=1$ \cite{Arkani-Hamed:2020gyp}, or if $\Delta_{1},\Delta_{2}$
lie on the principal series we may indeed have a singularity at $\epsilon=0$.
In order to avoid such complication we can consider generic weights
$\Delta_{1},\Delta_{2}$, together with assuming a soft UV behaviour
of the correlation functions. However, it would be interesting to
investigate a suitable contour prescription for the cases that yield
these poles.

\section{Pairings involving light-ray operators}\label{app:pairings}

We can now adapt the construction of Appendix \ref{app:shadows} for the case of the
light-ray transform, which requires certain additional care due to
the Lorentzian signature. This will give us explicit formulas for
the distributions $\langle LL\rangle$ and $\langle LO\rangle$ which
can then be used to contract our Lorentzian OPE block. Consider then
the definition (\ref{eq:ltre}) contracted with $O_{\Delta_{2}}^{m}$

\begin{align}
\langle L[O_{\Delta_{1}}^{m}](z_{1})O_{\Delta_{2}}^{m}(z_{2})\rangle & =\int_{z_{1}}^{z_{1}^{+}}\frac{dz_{P}}{z_{P1}^{2-\Delta_{1}}}\langle O_{\Delta_{1}}^{m}(z_{P})O_{\Delta_{2}}^{m}(z_{2})\rangle\,\,\,\,,\,\,\Delta_{1},\Delta_{2}\in1+i\mathbb{R}^{+}\nonumber \\
 & =\frac{D_{2}\delta(i\Delta_{1}-i\Delta_{2})}{(\Delta_{1}-1)\bar{z}_{12}^{\Delta_{1}}}\int_{z_{1}}^{z_{1}^{+}}\frac{dz_{P}}{z_{P1}^{2-\Delta_{1}}(z_{P2}+i\epsilon)^{\Delta_{1}}}\,,
\end{align}
where the $i\epsilon$ prescription arises from assuming $\bar{z}_{12}>0$.
The weights of the denominator make the integral conformally covariant.
Moreover it appears to have only a single singularity (a branch-cut),
hence the residue theorem shows that it vanishes. However, from the
analysis of the shadows in Appendix \ref{app:shadows} we expect it to be proportional
to $\delta(z_{12})$. To see how, introduce again a regulator $\delta\to0$,
together with the extra puncture $z_{3}\to\infty$, and unfix the
gauge:

\begin{align}
\int_{z_{1}}^{z_{1}^{+}}\frac{dz_{P}}{z_{P1}^{2-\Delta_{1}}(z_{P2}+i\epsilon)^{\Delta_{1}}} & =\frac{1}{z_{21}^{1+\delta}}\int_{z_{1}}^{z_{1}^{+}}\lim_{z_{3}\to\infty}\frac{dz_{P}z_{21}^{1+\delta}z_{31}^{1-\delta-\Delta_{1}}z_{32}^{\Delta_{1}-1}}{z_{P1}^{2-\Delta_{1}}(z_{P2}+i\epsilon)^{\Delta_{1}+\delta}(z_{3P}+i\epsilon)^{-\delta}}\nonumber \\
 & =\frac{1}{z_{21}^{1+\delta}}\int_{z_{1}}^{z_{1}^{+}}\frac{dz_{P}z_{12}^{1+\delta}z_{31}^{1-\delta-\Delta_{1}}z_{32}^{\Delta_{1}-1}}{z_{P1}^{2-\Delta_{1}}(z_{P2}+i\epsilon)^{\Delta_{1}+\delta}(z_{3P}+i\epsilon)^{-\delta}}\nonumber \\
 & =\frac{2i\sin\pi(\Delta_{1}+\delta)}{z_{21}^{1+\delta}}\int_{-\infty}^{0}\frac{dt}{(-t)^{\Delta_{1}+\delta}(1-t)^{-\delta}}\nonumber \\
 & =2i\sin\pi\Delta_{1}\frac{\Gamma(\Delta_{1}-1)\Gamma(1-\Delta_{1})}{\Gamma(-\delta)z_{21}^{1+\delta}}\nonumber \\
 & =\frac{4\pi i}{\Delta_{1}-1}\delta(z_{21})\,.
\end{align}
In the first step is crucial that $z_{1}$ is the origin of the Poincare
patch but it is not gauge fixed to $z_{1}=-\infty$ until the third
step (we would otherwise be studying the singular configuration $z_{13}\to0$).
In the third step we have proceeded as in section 2 and closed the
contour on either of the cuts. Finally we have used the reflection
formula together with the representation $-\delta/z_{21}^{1+\delta}\to\delta(z_{21})$
for integration regions with $z_{2}>z_{1}$, see e.g. \cite{Guevara:2019ypd}.
The result is then \footnote{In the same way as the results in \cite{Kravchuk:2018htv}, this kernel should be regarded as a distributional pairing obtained from a scalar two-point function and not as a time ordered two-point function itself (which vanishes since light-ray operators anihilate the vacuum). }

\begin{equation}
\langle L[O_{\Delta_{1}}^{m}](z_{1})O_{\Delta_{2}}^{m}(z_{2})\rangle=\frac{2\pi iD_{2}}{(\Delta_{1}-1)^{2}}\delta(i\Delta_{1}-i\Delta_{2})\times\frac{\delta(z_{12})}{\bar{z}_{12}^{\Delta_{1}}}\,\,\,,\Delta_{1},\Delta_{2}\in1+i\mathbb{R}^{+}\,.\label{eq:LOp}
\end{equation}

We can now further compute the $\langle LL\rangle$ pairing by light-transforming
the second operator

\begin{align}
\langle L[O_{\Delta_{1}}^{m}](z_{1})L[O_{\Delta_{2}}^{m}](z_{2})\rangle & =\int_{z_{2}}^{z_{2}^{+}}\frac{dz_{P}}{z_{P2}^{2-\Delta_{1}}}\langle L[O_{\Delta_{1}}^{m}](z_{1})O_{\Delta_{2}}^{m}(z_{P})\rangle\nonumber \\
 & =\frac{2\pi iD_{2}}{(\Delta_{1}-1)^{2}}\frac{\delta(i\Delta_{1}-i\Delta_{2})}{\bar{z}_{12}^{\Delta_{1}}}\int_{z_{2}}^{z_{2}^{+}}\frac{dz_{P}}{z_{P2}^{2-\Delta_{1}}}\delta(z_{P1})\,,
\end{align}
where we further assumed $z_{1}>z_{2}$ (besides our previous condition
$\bar{z}_{1}>\bar{z}_{2})$. The function can then be analytically
continued for generic configurations:

\begin{equation}
\langle L[O_{\Delta_{1}}^{m}](z_{1})L[O_{\Delta_{2}}^{m}](z_{2})\rangle=\frac{2\pi iD_{2}}{(\Delta_{1}-1)^{2}}\frac{\delta(i\Delta_{1}-i\Delta_{2})}{\bar{z}_{12}^{\Delta_{1}}z_{12}^{2-\Delta_{1}}}\,.\label{eq:llc}
\end{equation}

As an application of (\ref{eq:LOp}), let us compute the 3-point function
by contracting the full Lorentzian block (\ref{eq:lbck}). We obtain

\begin{align}
\langle O_{\Delta}(z_{1})O_{\Delta}(z_{2})L[O_{\Delta_{3}}^{m}]\rangle & =\int_{0}^{i\infty}\frac{d\nu}{2\pi i}\frac{K(\nu)}{z_{21}^{\Delta}\bar{z}_{21}^{\Delta}}\int_{\mathbb{R}}\frac{d\bar{z}_{P}\bar{z}_{21}^{1-h}}{\bar{z}_{p1}^{1-h}\bar{z}_{p2}^{1-h}}\int_{\mathbb{R}}\frac{dz_{P}z_{21}^{1-h}}{z_{p1}^{1-h}z_{p2}^{1-h}}\langle O_{2h}^{m}(z_{p},\bar{z}_{P})L[O_{\Delta_{3}}^{m}]\rangle\nonumber \\
 & =i\frac{D_{2}K(\nu)}{z_{21}^{\Delta}\bar{z}_{21}^{\Delta}(\Delta_{3}-1)^{2}}\int_{\mathbb{R}}\frac{d\bar{z}_{P}\bar{z}_{21}^{1-h}}{\bar{z}_{p1}^{1-h}\bar{z}_{p2}^{1-h}\bar{z}_{p3}^{2h}}\int_{\mathbb{R}}\frac{dz_{P}z_{21}^{1-h}}{z_{p1}^{1-h}z_{p2}^{1-h}}\delta(z_{p3})\nonumber \\
 & =i\frac{D_{2}K(\nu)}{z_{21}^{\Delta}\bar{z}_{21}^{\Delta}(\Delta_{3}-1)^{2}}\frac{z_{21}^{1-h}}{z_{13}^{1-h}z_{23}^{1-h}}\frac{\bar{z}_{21}^{h}}{\bar{z}_{13}^{h}\bar{z}_{23}^{h}}e^{i\pi (h-1)}\int_{\mathbb{R}}\frac{d\bar{z}_{P}\bar{z}_{21}^{1-2h}\bar{z}_{13}^{h}\bar{z}_{23}^{h}}{\bar{z}_{1p}^{1-h}\bar{z}_{2p}^{1-h}\bar{z}_{p3}^{2h}}\nonumber \\
 & =i\frac{D_{2}K(\nu)}{z_{21}^{\Delta}\bar{z}_{21}^{\Delta}(\Delta_{3}-1)^{2}}\frac{z_{21}^{1-h}}{z_{13}^{1-h}z_{23}^{1-h}}\frac{\bar{z}_{21}^{h}}{\bar{z}_{13}^{h}\bar{z}_{23}^{h}}e^{i\pi (h-1)} \times2\pi i\frac{\Gamma(1-2h)}{\Gamma(1-h)^{2}}\nonumber \\
 & =\frac{2\pi iD_{3}}{(1-\Delta_3)}\frac{e^{i\pi (h-1)}e^{-i\pi h }}{z_{21}^{\Delta+h-1}\bar{z}_{21}^{\Delta-h}z_{13}^{1-h}z_{23}^{1-h}\bar{z}_{31}^{h}\bar{z}_{23}^{h}}\,,\label{eq:3ptfin}
\end{align}
where the $\bar{z}_{P}$ integral was computed in (\ref{eq:sndf})
with appropriate $i\epsilon$ prescriptions. We see this matches the
correlation function (\ref{eq:3ptlightf}).

\bibliography{softoperators.bib}
\bibliographystyle{utphys}

\end{document}